\documentclass[journal]{IEEEtran}

\usepackage{setspace,cite}

\usepackage{multirow,multicol}
\usepackage[table]{xcolor}

\usepackage[tbtags]{amsmath}
\usepackage{amsbsy}
\usepackage{amssymb}
\usepackage{amsfonts}
\usepackage{amsthm}

\usepackage{graphicx}
\usepackage{algorithmic}
\usepackage{algorithm}
\usepackage{balance}
\usepackage{rotating}
\usepackage{multirow}
\usepackage{subcaption}
\usepackage{fancyvrb}
\usepackage{latexsym}
\usepackage{verbatim}
\usepackage{esint}

\usepackage[pagebackref=true,breaklinks=true,letterpaper=true,colorlinks,bookmarks=false]{hyperref}

{\begin{list}               
    {$\bullet$ \hfill}{
        \setlength{\leftmargin}{\parindent}
        \setlength{\parsep}{0.04\baselineskip}
        \setlength{\itemsep}{0.5\parsep}
        \setlength{\labelwidth}{\leftmargin}
        \setlength{\labelsep}{0em}}
    }
{\end{list}}

\providecommand{\eref}[1]{\eqref{#1}}  
\providecommand{\cref}[1]{Chapter~\ref{#1}}

\providecommand{\fref}[1]{Figure~\ref{#1}}

\providecommand{\R}{\ensuremath{\mathbb{R}}}

\providecommand{\E}{\ensuremath{\mathbb{E}}}

\providecommand{\norm}[1]{\lVert#1\rVert}

\providecommand{\bydef}{\overset{\text{def}}{=}}

\renewcommand{\vec}[1]{\ensuremath{\boldsymbol{#1}}}

\providecommand{\calD}{\mathcal{D}}
\providecommand{\calE}{\mathcal{E}}

\providecommand{\va}{\vec{a}}

\providecommand{\vf}{\vec{f}}

\providecommand{\vh}{\vec{h}}

\providecommand{\vr}{\vec{r}}

\providecommand{\vu}{\vec{u}}

\providecommand{\vw}{\vec{w}}

\providecommand{\vy}{\vec{y}}
\providecommand{\vz}{\vec{z}}

\providecommand{\valpha}{\vec{\alpha}}

\providecommand{\vyhat}{\boldsymbol{\widehat{y}}}

\newcommand{\argmin}[1]{\mathop{\underset{#1}{\mbox{argmin}}}}

\newcommand\nwidth{0.14}
\newcommand\nspace{-1}
\newcommand\nspacetwo{0}

\begin{document}

\title{Image Reconstruction of Static and Dynamic Scenes through Anisoplanatic Turbulence}

\author{Zhiyuan~Mao,~\IEEEmembership{Student~Member,~IEEE,}
        Nicholas~Chimitt,~\IEEEmembership{Student~Member,~IEEE,}
        and~Stanley~H.~Chan,~\IEEEmembership{Senior~Member,~IEEE}
\thanks{The authors are with the Department
of Electrical and Computer Engineering, Purdue University, West Lafayette,
IN, 47906 USA. Email: \{mao114, nchimitt, stanchan\}@purdue.edu.}}

\maketitle

\begin{abstract}
Ground based long-range passive imaging systems often suffer from degraded image quality due to a turbulent atmosphere. While methods exist for removing such turbulent distortions, many are limited to static sequences which cannot be extended to dynamic scenes. In addition, the physics of the turbulence is often not integrated into the image reconstruction algorithms, making the physics foundations of the methods weak. In this paper, we present a unified method for atmospheric turbulence mitigation in both static and dynamic sequences. We are able to achieve better results compared to existing methods by utilizing (i) a novel space-time non-local averaging method to construct a reliable reference frame, (ii) a geometric consistency and a sharpness metric to generate the lucky frame, (iii) a physics-constrained prior model of the point spread function for blind deconvolution. Experimental results based on synthetic and real long-range turbulence sequences validate the performance of the proposed method.
\end{abstract}


%
\IEEEpeerreviewmaketitle

\section{Introduction}
Ground-based long-range passive imaging systems often suffer from degraded image quality due to a turbulent atmosphere. The cause of such degradation is complicated, as the atmospheric turbulence is affected by temperature, wind velocity, humidity, air pressure, and many other factors \cite{Tatarskii1961}. From the image quality point of view, the influence of turbulence on an image is often demonstrated through the random warping and blurring of the image. If the distortions are spatially and temporally varying, we refer to the turbulence as \emph{anisoplanatic} \cite{roggemann1996imaging, Fried82}. Anisoplanatic turbulence is common in ground-to-ground systems where the object distance is long and the field of view is large.  

Image processing techniques for recovering images from anisoplanatic turbulence have been studied for decades. However, while the literature is rich, existing methods have several limitations: (i) Many methods are designed for the easier case of isoplanatic turbulence, i.e., the blur is spatially invarying.  Methods for anisoplanatic turbulence are considerably fewer, among which most are designed for static scenes only. Moving objects remains very challenging. (ii) There is generally a disconnect between the reconstruction algorithm and the physics of the turbulence. Some existing image processing methods use over-simplified models that are not justified and explainable by physics. (iii) Existing benchmark evaluation sequences are often collected by short-range hot-air burners. The turbulence in these sequences is usually isoplanatic. The long-range anisoplanatic effects are not captured.

The goal of this paper is to present a robust image reconstruction algorithm for anisoplanatic turbulence. Our method applies to both static and dynamic scenes and is evaluated using a comprehensive set of real and synthetic data. \fref{fig:pipeline} summarizes the three contributions of this paper:

\begin{figure*}[!]
\centering
\includegraphics[width=\linewidth]{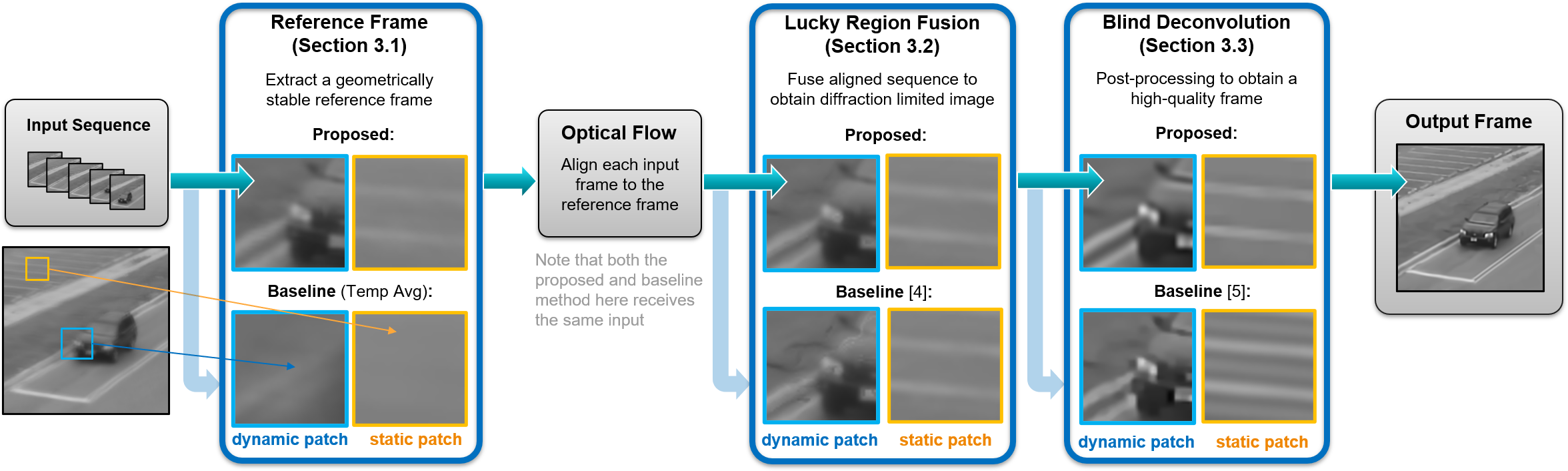}
\caption{\textbf{Contributions of this paper}. (1) We use space-time non-local averaging to construct a reference frame for optical flow. Our reference frame can maintain the moving objects while suppressing turbulence. The baseline method using temporal averaging tends to wash out the objects. (2) We propose new metrics to extract the lucky regions from the optical warped images. Compared to a baseline method \cite{Milanfar2013}, our method is more robust to registration error near the moving object. (3) We propose a new blind deconvolution algorithm based on Zernike decomposition. Compared to the baseline method \cite{Shan2008}, our method generates fewer artifacts.}
\label{fig:pipeline}
\end{figure*}

\begin{enumerate}
    \item We propose a new method to construct a reliable reference frame. Our method is based on a non-local space-time averaging concept. Compared to existing approaches that are based on temporal averaging, the new method is able to preserve the moving content while stabilizing the turbulent background. (See Section 3.A.)
    \item We propose a new method to generate the lucky frame by combining two image quality metrics: A geometric consistency metric and a sharpness metric. Compared to existing lucky fusion methods that generate artifacts, our method offers significantly better object consistency. (See Section 3.B.)
    \item We propose a new prior for the point spread functions (PSFs) to improve the blind deconvolution step. Our prior is based on a linear expansion model of the PSFs through simulation in the Zernike space. Compared to generic blind deconvolution methods, our method offers better reconstruction quality and robustness. (See Section 3.C.)
\end{enumerate}

Besides presenting the ideas, this paper also provides detailed experimental results using synthetic data and real data. For real data, we do not only compare with the commonly used short-range hot-air turbulence sequences but also long-range sequences. These results will be shown in Section 4. 

This paper focuses on incoherent imaging. By incoherence we meant that no active light sources are used to illuminate the scene. The alternative to incoherent imaging is coherent imaging, e.g., using adaptive optics \cite{tyson2010principles} and speckle imaging techniques \cite{Labeyrie1970,Knox1974,Lohmann1983,Carrano2002}. Coherent imaging requires dedicated instrumentation, e.g., telescopes in astronomy applications, which are not always feasible for systems constrained by size, weight, and power.

\section{Background}
As light propagates through a turbulent medium, the wavefront is distorted and so the images are warped and blurred. In this section, we provide a brief overview of the image degradation process and explain the origins of several mainstream image processing methods.

\subsection{The Optics of Imaging through Turbulence}
We first discuss the optics of how light propagates through turbulence. Much of the materials in this section can be found in textbooks such as \cite{roggemann1996imaging, Goodman_StatisticalOptics}. We highlight a few key concepts in order to explain our algorithm. 

\textbf{Statistics of the phase}. The optical distortion due to turbulence is caused by the inhomogeneous refractive index of the medium. As hot and cold air moves, the refractive index changes spatially and temporally. When a wave passes through the turbulence, the phase of the wavefront, $\phi$, will be distorted, accumulated, and propagated. The physics of such phase distortion is well known in the literature \cite[Section 8.3]{Goodman_StatisticalOptics}. It can be shown that the phase is best characterized by the \emph{structure function} \cite[Eq. 1.13]{Tatarskii1961} which is a variant of the autocorrelation function. By using the Kolmogorov spectrum \cite{Kolmogorov1941}, it follows that the structure function $\calD_{\phi}$, assuming wide sense stationarity and spherical symmetry, is \cite[Eq. 5.3]{Fried66optical}
\begin{equation}
    \calD_{\phi}(|\vf - \vf'|) = 6.88(|\vf - \vf'|/r_0)^{5/3},
    \label{eq: structure function}
\end{equation}
where $\vf$ and $\vf'$ are the (2D) spatial frequencies of the phase, and $r_0$ is called the atmospheric coherence diameter (or the Fried parameter). The Fried parameter measures how strong the turbulence is. It is a function of the refractive index structure parameter $C_n^2$, the path length $L$, and the wave number $k^2$ \cite[Eq. 3.67]{roggemann1996imaging}. 

\begin{figure*}[!t]
	\centering
	\begin{tabular}{cc}
    	\includegraphics[height=4cm]{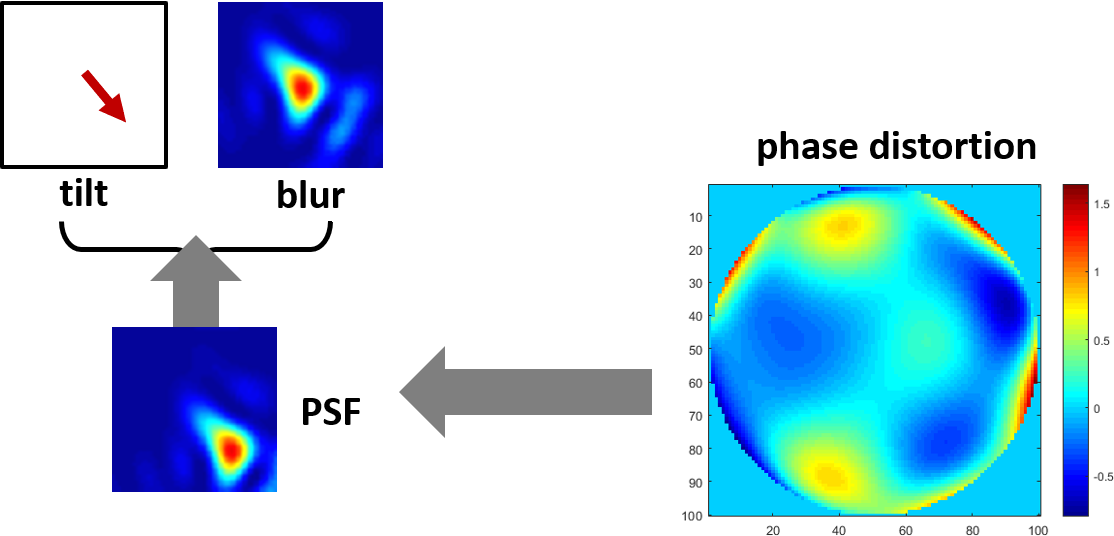} & 
    	\includegraphics[height=4cm]{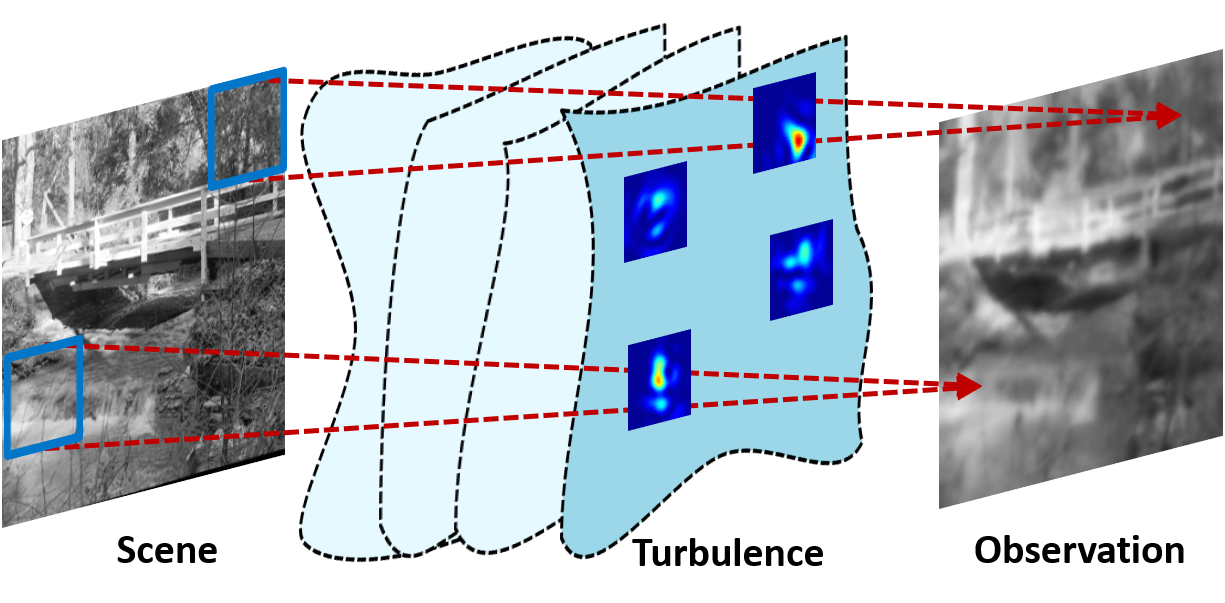} \\
        (a) Phase Distortion &  (b) Wave Propagation through Turbulence
    \end{tabular}
	\caption{\textbf{Image formation process in a turbulent medium}. (a) The origin of the turbulence distortion is the random phase caused by the changing refractive index. The phase will lead to a point spread function (PSF), which consists of tilt and blur. The tilt shifts the pixels, whereas the blur degrades the image. (b) As light travels from the object plane to the image plane, the PSFs are applied to individual blocks of pixels to generate a spatially-varying distortion. 
	}
	\label{fig:schematic}
\end{figure*}

\textbf{Optical transfer function}. The randomness of the phase distortion is translated to the images through the point spread function (PSF) or the optical transfer function (OTF), where OTF is the Fourier transform of PSF. The OTF of turbulence is a product of two components:
\begin{equation}
H(\vf) = H_{\mathrm{atm}}^{\phi}(\vf) H_{\mathrm{dif}}(\vf),
\end{equation}
where the first OTF $H_{\mathrm{atm}}^{\phi}$ is caused by the random atmosphere, and the second OTF $H_{\mathrm{dif}}$ is caused by the fixed lens \cite[Eq. 3.17]{Fried66optical}. $H_{\mathrm{atm}}^{\phi}$ accounts for the random tilts and spatially varying blur, whereas $H_{\mathrm{dif}}$ accounts for the optics which is spatially invariant. The PSF of a simulated turbulence is illustrated in \fref{fig:schematic}, where we emphasize the decoupling of the tilt and the blur.

\textbf{Long and short exposures}. There are two important observations of $H_{\mathrm{atm}}^{\phi}$. First, $H_{\mathrm{atm}}^{\phi}$ is spatially varying because $\phi$ changes from pixel to pixel. If the field of view is within the isoplanatic angle, then $\phi$ is not random and so 
$H_{\mathrm{atm}}^{\phi}$ will become spatially invarying \cite[Ch. 4]{roggemann1996imaging}.

Second, the exact expression of $H_{\mathrm{atm}}^{\phi}(\vf)$ is can be determined using convolutions \cite[Eq. 2.44]{roggemann1996imaging}. Plugging the structure function $\calD_{\phi}$ into the equation, one can show that the statistical average of $H_{\mathrm{atm}}^{\phi}(\vf)$, known as the \emph{long-exposure} OTF, is \cite[Eq. 3.16]{Fried66optical}:
\begin{equation}
\hspace{-2ex} H_{\mathrm{LE}}(\vf) \bydef \E\left[ H_{\mathrm{atm}}^{\phi}(\vf) \right] = \exp\left\{-3.44\left(\frac{\lambda d |\vf|}{r_0}\right)^{5/3}\right\},
\label{eq: HLE}
\end{equation}
where $d$ is the focal length of the lens, $\lambda$ is the wavelength, and $r_0$ is the Fried parameter. Note that $H_{\mathrm{LE}}(\vf)$ is spatially \emph{invariant} since it is the average of a random process. 

In the turbulence literature, people also consider the statistical average of the OTF when the tilts in the phase distortions are removed. In this case, the tilt-free phase distortion $\varphi$ will lead to a different OTF, called the short-exposure OTF, defined as \cite[Eq. 5.9a]{Fried66optical}
\begin{align}
&H_{\mathrm{SE}}(\vf)
\bydef \E\left[ H_{\mathrm{atm}}^{\varphi}(\vf) \right] \notag \\
&= \exp\left\{-3.44\left(\frac{\lambda d |\vf|}{r_0}\right)^{5/3} \left[1-\left(\frac{\lambda d |\vf|}{D}\right)^{1/3}\right] \right\} \label{eq: HSE},
\end{align}
where $D$ is the aperture diameter. The difference between the long-exposure and the short-exposure is that the absence of tilts in the short-exposure equation. Therefore, the blur associated with $H_{\mathrm{SE}}$ is generally much weaker than that of $H_{\mathrm{LE}}$. This suggests that if we remove the tilts and then take am average, then the resulting deblurring problem will be easier to solve than the long-exposure case.

\textbf{Summary}. To summarize, the image formation process in a turbulent medium is originated from the phase. The phase causes tilt and blur, which then leads to a pixel-wise PSF, as shown in \fref{fig:schematic}(a). As light travels from the object plane to the image plane, the spatially varying PSFs will interact with the pixels in the object plane to create the distortion effect, as shown in \fref{fig:schematic}(b). The analysis explains the principle behind many image processing algorithms for turbulence. The idea is to remove the tilt, construct the short exposure expectation by taking an average, and then deblur. Our proposed method is also based on these steps.

\subsection{Image Reconstruction and Current Limitations}
Having discussed the image formation process, we now summarize the typical approaches in existing image reconstruction methods and comment on their limitations. 

\textbf{Step 1. Removing tilts}. The first step of a reconstruction method is to mimic the short-exposure procedure by removing the tilts. This is often done using the optical flow \cite{Liu2009optflow}. However, since all input frames of a turbulence-distorted sequence are randomly warped, we need to first construct a geometrically stable reference frame.

When the scene is static, one can take the temporal average as the reference frame like  \cite{aubaillyLucky, Lou2013, Milanfar2013, Anantrasirichai2013, Gilles2016, Hardie2017}. More advanced approaches such as robust principal component analysis (RPCA) can be used to extract the low-rank part as the reference frame. However, RPCA generally has comparable performance to temporal averaging as shown in section IV. B. 

Improvement of the reference frame extraction methods exists, e.g., the variational method by Xie et al. \cite{Xie2016}, and frame selection schemes by Anantrasirichai et al. \cite{Anantrasirichai2013} and Lau et al. \cite{Lau2017}. However, these techniques are designed for static scenes. Images containing moving objects still cannot be recovered. 

Reference extraction methods for moving scenes are available but scattered. Halder et al. \cite{Halder2015} assume that a certain segment of the video sequence does not contain any moving object so that a reference can be extracted. Oreifej et al. \cite{Oreifej2013} and Anantrasirichai et al. \cite{Anan2018} use background segmentation to separate the moving object from the static background. However, since turbulence-distorted images are usually blurry, the object boundaries of the segmentation are difficult to be determined precisely which can be seen from \fref{fig:movexp} in the experiment section.

Our proposed method does not require any segmentation. It is based on a space-time non-local averaging which can stabilize background while preserving moving objects.

\textbf{Step 2. Lucky image fusion}. While optical flow can align the pixels so that we can approximate the short-exposure frames, a simple averaging over all these short-exposure frames is problematic because the optical flow algorithms are not perfect. In addition, it is known in turbulence literature that a sharp region will occasionally appear because the random phase distortion is zero-mean \cite{Fried78}. Therefore, the most reasonable step here is to compare the spatial blocks over an extended period of frames, and pick the sharpest block. This step is known as the lucky image fusion. 

The goal of the lucky fusion step is to combine sharp image blocks to form a diffraction limited image. The fusion can be done in the wavelet domain as proposed by Anantrasirichai et al. \cite{Anantrasirichai2013}, or in the spatial domain by computing the local gradient, variance or distortions as proposed by Aubailly et al. \cite{aubaillyLucky}, Zhu and Milanfar \cite{Milanfar2013}, and Xie et al. \cite{Xie2016}. For robust PCA approaches such as Lau et al. \cite{Lau2017} and He et al. \cite{He2016}, the lucky image fusion step is done by refining the sparse components and adding them to the low-rank parts of the image.

\begin{figure*}[!]
	\centering
	\begin{tabular}{cc}
    	\includegraphics[width=0.45\linewidth]{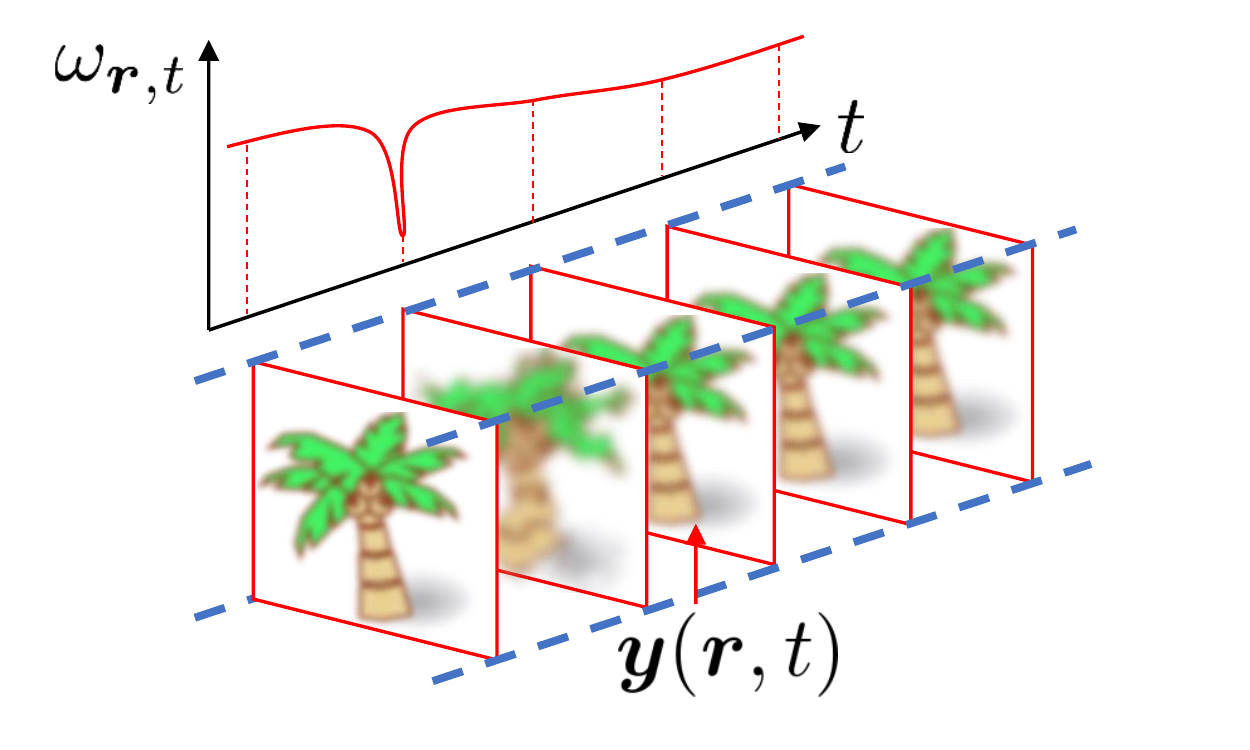} & 
    	\includegraphics[width=0.45\linewidth]{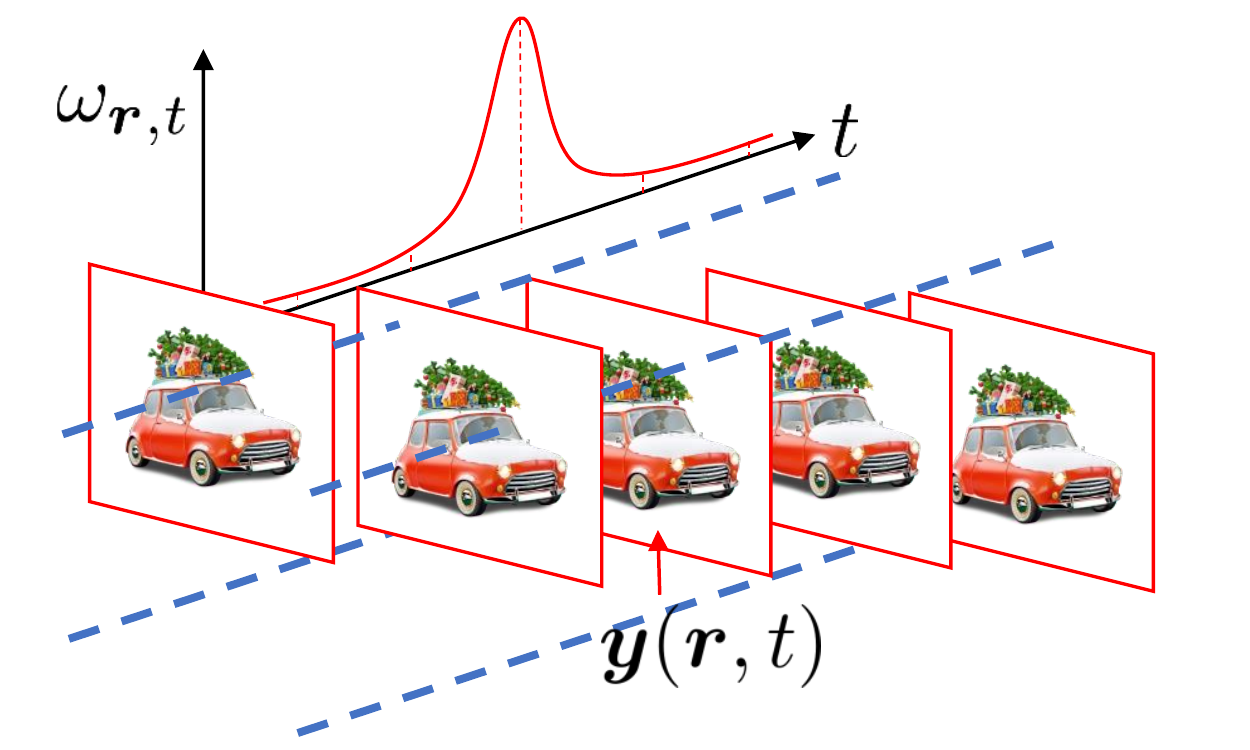} \\
        (a) Static Patches &  (b) Motion Patches
    \end{tabular}
	\caption{\textbf{How the proposed method in Section 3.A can handle both static and dynamic scenes}. (a) For static scene, jittering patches with similar weight are averaged while bad patches with significant blur or geometric distortion are rejected by a much smaller weight. (b) For dynamic scene, the motion patches will have large weights only at its neighbors. Thus, the weighted average will not wash out the object.
	}
	\label{fig:ref_illustration}
	\vspace{-2ex}
\end{figure*}

In our proposed method, the lucky image fusion is done by a new image quality metric that measures both the sharpness and consistency of the patches. The sharpness metric helps extracting the sharp patches from the sequence, whereas the consistency metric improves the robustness by rejecting any residual jittering pixels from the previous steps.

\textbf{Step 3. Blind deconvolution}. After the lucky region fusion, the residual blur that is remaining in the image is largely caused by the diffraction limit of the lens, plus some minor effects due to any averaging process and artifacts of the previous steps. Therefore, a blind deconvolution algorithm is applied at last to remove these residual blurs. 

Many turbulence mitigation methods, e.g. \cite{Milanfar2013, Anantrasirichai2013, He2016, Xie2016, Lau2017}, use off-the-shelf blind deconvolution algorithms such as the classical method by Shan et al. \cite{Shan2008}. However, the priors they used to define the point spread function are generic, e.g., $\ell_1$ or total variation. These priors have almost no connection with the statistical behavior of turbulence, in particular the 5/3-power kernel proved by Fried \cite{Fried66}. A more recent approach by Nieuwenhuizen et al. uses deep neural networks \cite{Nieuwenhuizen2019}. However, with limited publicly available datasets to train the network,  generalization of the method is limited. A physically more principled approach is by Hardie et al. \cite{Hardie2017}, where they construct the ideal short-exposure point spread function by estimating the turbulence parameters. However, since the optical flow is not perfect and the lucky fusion sometimes introduces extra blur \cite{Milanfar2013,aubaillyLucky}, the residual errors can contaminate the predicted ideal PSF in \cite{Hardie2017}. 

Our proposed method is to construct the prior via basis expansion, similar to the analysis of Fried \cite{Fried78} but straightly in the spatial domain. It is a semi-supervised learning integrating numerical simulation and sparse modeling.

\section{Proposed Method}
The proposed method consists of three ideas, each targeting at a specific step outlined in \fref{fig:pipeline}. In this section we present the details of these ideas.

\subsection{Space-time Non-Local Averaging}
As shown in \fref{fig:pipeline}, the first step of the turbulence mitigation process is to remove the tilts by using an optical flow algorithm. However, since the optical flow algorithm requires a reference image which is not available from the raw data, we need to construct it. Existing reference frame generation methods are largely based on temporal averaging. These methods fail to handle moving objects. Our goal is to develop a technique that can generate geometrically stable reference frames for both static and dynamic scenes.

\textbf{Proposed approach}.  We propose to consider two criteria for the reference frame. First, the reference image needs to keep any moving objects intact. A temporal average is not a good option here because the moving objects will be washed out. Second, the reference image needs to remove the pixel jittering caused by turbulence. This ensures that the reference frame is sufficiently stabilized for the optical flow algorithm. 

Let $\vy(\vr,t) \in \R^d$ be a $d$-dimensional patch located at pixel $\vr = (i,j)$ and time $t$. Surrounding the patch we create a spatial-temporal search window $\Omega_{\vr} \times \Omega_t$ such that the possible patches contained are $\{\vy(\vr+\Delta \vr, t + \Delta t) \;|\; \Delta \vr \in \Omega_{\vr}, \; \Delta t \in \Omega_t\}$. For each patch within this window, we define a pairwise distance with $\vy(\vr,t)$. This gives an array of distances measured at each space-time index $(\vr,t)$:
\begin{equation}
    \delta_{\vr,t}(\Delta \vr, \Delta t) = \left\| \vy(\vr,t) - \vy(\vr+\Delta \vr, t + \Delta t) \right\|^2.
\end{equation}

Now, for every $\Delta t$ (i.e., every frame in the temporal window) we scan through the spatial search window $\Omega_{\vr}$ and pick the $\Delta \vr$ which minimizes the distance:
\begin{equation}
    \overline{\delta}_{\vr,t}(\Delta t) = \underset{\Delta \vr}{\min} \;\; \delta_{\vr,t}(\Delta \vr, \Delta t).
    \label{eq: D distance}
\end{equation}
Consequently, we construct a space-time non-local weight
\begin{equation}
    w_{\vr,t}(\Delta t) = \exp\{-\beta \overline{\delta}_{\vr,t}(\Delta t)\},
    \label{eq: ref weight}
\end{equation}
which is a function of the frame index $\Delta t$, and $\beta$ is a hyper-parameter controlling the decay rate of the weight. The estimated reference frame is the weighted average:
\begin{equation}
    \vyhat_{\text{ref}}(\vr,t) = \frac{\sum\limits_{\Delta t \in \Omega_t } w_{\vr,t}(\Delta t) \vy(\vr,t+\Delta t)}{\sum\limits_{\Delta t \in \Omega_t} w_{\vr,t}(\Delta t)}.
    \label{eq: weighted average}
\end{equation}
For overlapping patches, we take the average of the overlapping spatial locations to construct the final reference image.

\textbf{Intuition}. The intuition of the proposed method can be seen from the illustration shown in \fref{fig:ref_illustration}. In this figure we consider two cases: A patch of a static object, and a patch of a moving object. 

When the patch is static, the distortion caused by turbulence will be mostly random jittering. The jittering is a random variable with zero mean. Therefore, if we take the current patch $\vy(\vr,t)$ and search for its neighbors in the adjacent frames, we are likely to find many neighbors. This can be seen from the high (and steady) values of the weights in \fref{fig:ref_illustration}(a), except some severely jittered patches which give small weights. The usage of the minimum operation is critical when defining $\overline{\delta}_{\vr,t}(\Delta t)$. If the minimum-distant neighboring patch (measured in terms of the $\ell_2$ distance) still does not match well with the current patch, then this neighboring patch is likely to be either severely distorted or is a moving patch. In this case $w_{\vr,t}(\Delta t)$ will be small. 

When a patch contains moving objects, the distance $\delta_{\vr,t}(\Delta\vr, \Delta t)$ will be large for all $(\Delta\vr, \Delta t)$ because we are not able to find a good match within the spatial-temporal window. If this happens, $\overline{\delta}_{\vr,t}(\Delta t)$ will remain large and so the weight $w_{\vr,r}(\Delta t)$ will be small. As illustrated in \fref{fig:ref_illustration}(b), the weight is large at the current time index, and is small for other time indices. By computing the weighted average, the moving object will remain intact because we do not perturb the current patch. 

\textbf{Window size.} In principle, the spatial window size should be chosen according to the strength of the turbulence. However, the strength of the turbulence is not observable and any values obtained are essentially estimates, we take a different approach of choosing the window size empirically. Through our experimentation, we found that the window size of 11 pixels generally leads to satisfactory reconstruction results. 

\textbf{Hyper-parameter $\beta$}. The hyper-parameter $\beta$ in \eref{eq: ref weight} is chosen according to the turbulence strength. Turbulence strength is usually quantified by the aperture-to-coherence ratio $D/r_0$. Here, $D$ is the aperture diameter and $r_0$ is the Fried parameter \cite{roggemann1996imaging}. A typical value of $D/r_0$ in incoherent imaging ranges from 1 to 5\cite{Fried78}, with 1 being weak turbulence and 5 being strong.

The parameter $\beta$ is empirically determined as follows. For a specific $D/r_0$ value, we synthetically generate random tilts to an image containing $64\times 64$ point sources using the tilt statistics (via the Zernike decompositions) \cite{Our_paper}. We choose to work with synthetic point sources because it is content independent. We sweep through a range of $\beta$'s in the space-time non-local averaging operation, and pick the $\beta$ that minimizes the error $\norm{\vyhat_\text{ref}(\vr,t) - \vy_{\text{true}}(\vr,t)}$, where $\vy_{\text{true}}$ is the ground truth image. 

\begin{figure}[h]
\centering
\includegraphics[width=\linewidth]{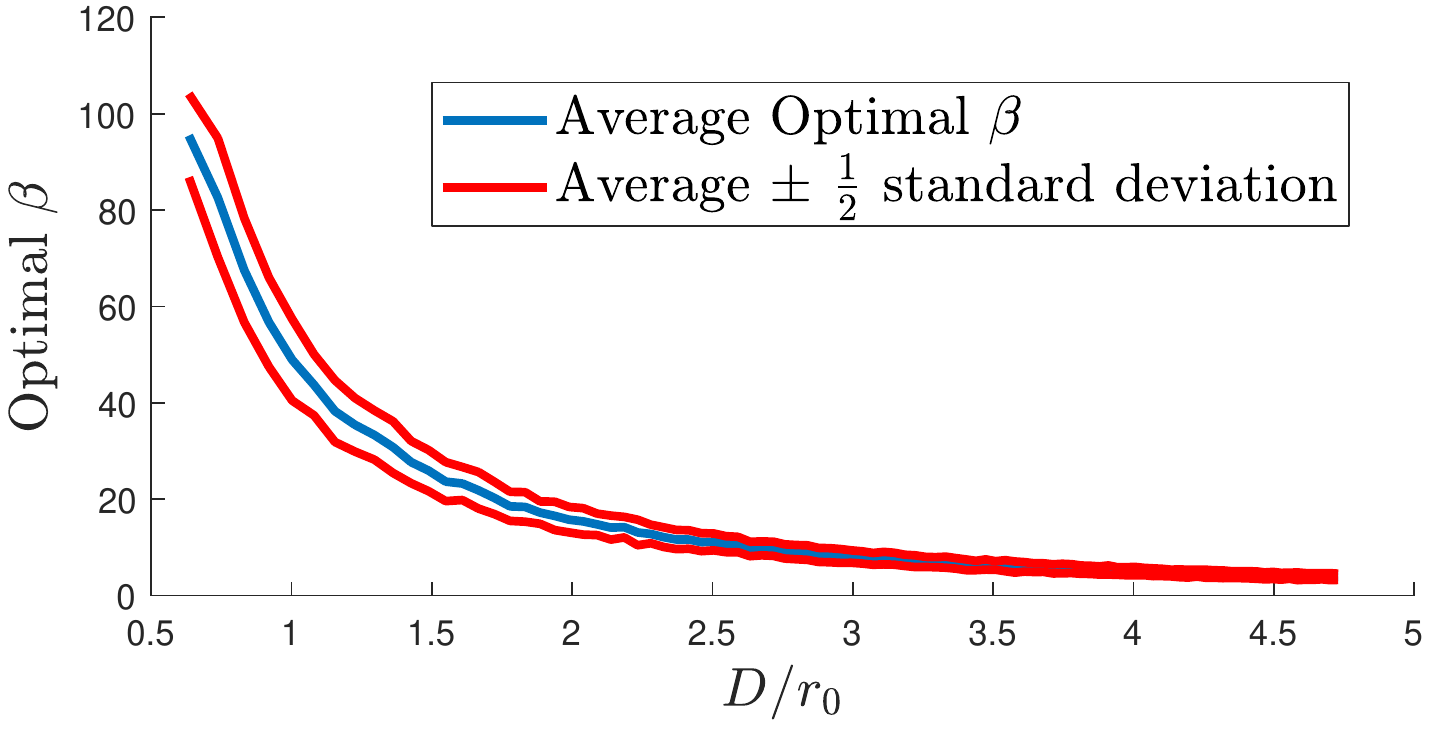}
\caption{\textbf{Hyper-parameter $\beta$}. We choose the optimal $\beta$ according to the turbulence strength $D/r_0$. The blue curve is averaged over 100 trials.}
\label{fig: beta}
\vspace{-2ex}
\end{figure}

The result of the experiment is shown in \fref{fig: beta}, where we plot the optimal $\beta$ as a function of $D/r_0$. The result gives an empirically optimal $\beta$, which suggests that as the turbulence becomes stronger (larger $D/r_0$), more frames are required to average out the randomness (hence $\beta$ drops). 

\subsection{Lucky Region Fusion}
The goal of the lucky region fusion is to extract sharp patches from the optical-flow aligned frames to form an image that is roughly diffraction limited. Conventional lucky region fusion techniques \cite{Milanfar2013,aubaillyLucky} achieve the goal by computing the sharpness of a patch and pick the one with the highest sharpness score along the time axis. However, the sharpest patch may not necessarily be consistent with its neighbors. The winner-take-all approach is also not ideal because it creates discontinuity along object boundaries.

\textbf{Proposed approach}. The proposed lucky region fusion includes two new ideas. First, we introduce a geometric consistency metric in addition to the sharpness metric used in existing approaches. The geometric consistency metric ensures that the sharp patch is not due to artifacts. The second idea is to replace the winner-take-all by a weighted average. The weighted average ensures that the reconstructed lucky image has consistent object boundaries.

Let $\vy_{\text{flow}}(\vr,t) \in \R^{d}$ be a $d$-dimensional patch of the $t$-th optical flow aligned image centered at pixel $\vr$, and let $\vyhat_{\text{ref}}(\vr,t) \in \R^{d}$ be the corresponding patch of the estimated reference frame. We define the geometric consistency between patches $\vy_{\text{flow}}(\vr,t)$ and $\vyhat_{\text{ref}}(\vr,t)$ as
\begin{align}
    &\delta^{\text{G}}(\vr,t) = \| \vy_{\text{flow}}(\vr,t) - \vyhat_{\text{ref}}(\vr,t) \|^2. \label{eq: delta G}
\end{align}
The term $\delta^{\text{G}}$ is called the geometric consistency, because it measures how close the aligned image is compared to the reference. If they are close, then the patch will be used to form the lucky fusion. The other metric measures the sharpness, defined as
\begin{equation}
\delta^{\text{S}}(\vr,t) = \|\nabla \vy_{\text{flow}}(\vr,t)\|_1, \label{eq: delta S}
\end{equation}
where $\nabla$ is the spatial gradient operator. If an image is sharp, then the gradient $\vy_{\text{flow}}(\vr,t)$ will have a large magnitude.

The overall impact of the two metrics is defined through the weight
\begin{align}
    w_{\vr,t}(\Delta t) &= 
    \underset{\text{geometric}}{\underbrace{\exp \left\{ -\alpha_1\delta^{\text{G}}(\vr,t+\Delta t) \right\}}} \notag \\
    & \quad \quad\quad \times 
    \underset{\text{sharpness}}{\underbrace{
    \exp\left\{ \alpha_2 \delta^{\text{S}}(\vr,t+\Delta t) \right\}}},
\end{align}
where $\alpha_1$ and $\alpha_2$ are two user defined parameters controlling the cutoff of the inliers and outliers. The typical values of $\alpha_1$ and $\alpha_2$ are determined empirically by verifying the percentile of the scores. One may increase the value of $\alpha_1$ if the optical flow output contains more errors. Given the weights, the lucky region patch at pixel $\vr$ and time $t$ is then
\begin{equation}
\vyhat_{\text{lucky}}(\vr,t) = \frac{\sum\limits_{\Delta t \in \Omega_t } w_{\vr,t}(\Delta t) \; \vy_{\text{flow}}(\vr,t+\Delta t)}{ \sum\limits_{\Delta t \in \Omega_t} w_{\vr,t}(\Delta t) }. 
\label{eq: lucky weight}
\end{equation}

\begin{figure}[t]
	\centering
	\includegraphics[width=\linewidth]{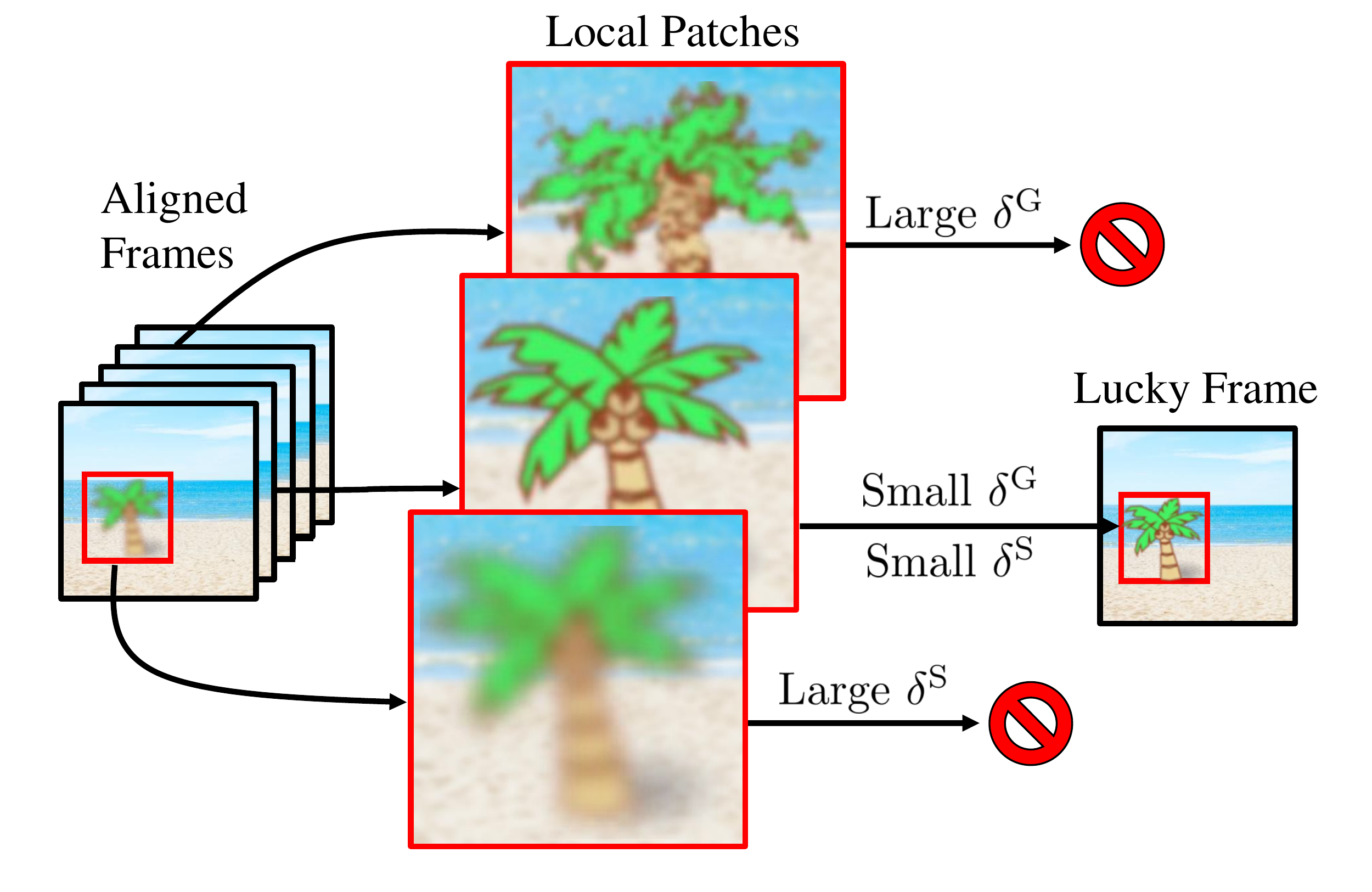}
	\caption{\textbf{Lucky region metrics}. The optical flow aligned frames are assumed to be motion compensated. When determining a lucky patch, the method checks two scores: (i) the geometric consistency score which asks the candidate patch to look similar to the reference, and (ii) the sharpness score which asks the candidate patch to be sharp. Weighted average is formed across all the optical flow aligned frames. }
	\label{fig:lucky}
\end{figure}

\begin{figure}[t]
	\centering
	\begin{tabular}{ccc}
	\hspace{-2ex}\includegraphics[width=0.32\linewidth]{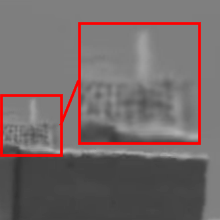}&
	\hspace{-2ex}\includegraphics[width=0.32\linewidth]{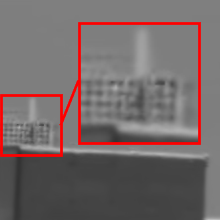}&
	\hspace{-2ex}\includegraphics[width=0.32\linewidth]{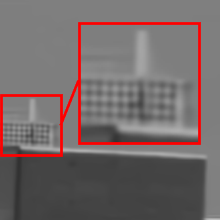}\\
	(a) Input & (b) w/o $\delta^{\text{G}}(\vr,t)$ & (c) w/ $\delta^{\text{G}}(\vr,t)$
	\end{tabular}
	\caption{\textbf{Effect of $\delta^{\text{G}}(\vr,t)$}. Results of lucky region fusion generated by using $\delta^{\text{G}}(\vr,t)$ and without $\delta^{\text{G}}(\vr,t)$. Note the improved stabilization of the edges, and geometric consistency.}
	\label{fig:similarity}
	\vspace{-2ex}
\end{figure}

\textbf{Intuition}. \fref{fig:lucky} illustrates the intuition of the weight $w_{\vr,t}(\Delta t)$. By virtue of optical flow, large motions are compensated. This means that the remaining distortions are mostly jitter. If the jittering in the patch $\vy_{\text{flow}}(\vr,t)$ is still strong, then $\delta^{\text{G}}(\vr,t)$ will be large because $\vy_{\text{flow}}(\vr,t)$ is different from $\vyhat_{\text{ref}}(\vr,t)$. However, if the jittering is weak, then the patch is reliable and we should use it. Thus, a candidate patch is accepted to be used in the weighted average only when it has little geometric warping and it is sharp.

In \fref{fig:similarity} we show a comparison between the lucky region fusion results generated with and without the geometric consistency term $\delta^{\text{G}}(\vr,t)$. (The necessity of the sharpness metric is clear in the literature, for example \cite{aubaillyLucky,Milanfar2013}.) As we can observe from \fref{fig:similarity}, the geometric consistency term helps rejecting patches that are sharp but contain jittering artifacts.

\subsection{Blind Deconvolution}

The objective of the blind deconvolution step is to remove the residual blur in the lucky frame. This blur comes from three sources: (i) The diffraction limit of the optical system, which is usually not severe. (ii) Weighted temporal averaging in the lucky region fusion step. (iii) Limited number of available frames, which does not allow us to pick the sharpest lucky frame. 

\textbf{Proposed approach}. Following most of the literature on blind deconvolution, our proposed method starts with a pair of alternating minimization steps. The first step update the latent image, and the second step updates the unknown point spread function (PSF):
\begin{align}
\vz^{k+1} &= \argmin{\vz}\; \|\vy - \vh^k \circledast \vz\|^2 + \lambda g(\vz), \label{eq: vx subproblem}\\
\vh^{k+1} &= \argmin{\vh}\; \|\vy - \vh \circledast \vz^{k+1} \|^2 + \gamma r(\vh). \label{eq: vv subproblem}
\end{align}
In this pair of equations, $\vy$ is the entire lucky image which strictly speaking should be denoted by $\vyhat_{\text{lucky}}(t)$. We dropped the time index because blind deconvolution is performed per frame. The variable $\vh$ is the unknown PSF, and $\vz$ is the latent clean image. The functions $g(\cdot)$ and $r(\cdot)$ are regularizations for $\vz$ and $\vh$, respectively. In this paper, we adopt the Plug-and-Play prior \cite{chan2017plug} for $g(\cdot)$ with BM3D as the denoiser. The regularization $r(\cdot)$ will be discussed next. Additional details of the Plug-and-Play prior can be found in \cite{Schniter_PnP_MRI, Buzzard_Chan_Sreehari_2017, Chan_2019}.

\textbf{Modeling $\vh$}. A major difference between a turbulence deconvolution and a standard deconvolution is that we have a well-defined model for the turbulence PSF as we discussed in the Background section. Specific to the context of turbulence, we highlight a few key points:
\begin{itemize}
    \item The turbulence PSF is defined through the random phase distortion function $\phi(\vf)$ in the frequency domain. See \cite[Eqs 6-9]{Our_paper}.
    \item The random phase distortion $\phi(\vf)$ can be represented in the Zernike space via \cite[Eq. 15]{Our_paper}:
    \begin{equation}
        \phi(\vf) = \sum_{j=1}^{N} a_j Z_j(\vf).
        \label{eq: Zernike}
    \end{equation}
    with $Z_j$ being the $j$-th Zernike Basis, and $a_j$ is the corresponding Zernike coefficient.
    \item The Zernike coefficients have physical interpretations. E.g., $a_2$ and $a_3$ are the tilts, and $a_4$ is the defocus, etc. The statistics of $a_j$'s are defined through intermodal and spatial correlations. See \cite[Eq 20, Eq 33]{Our_paper}. 
\end{itemize}
In order to encapsulate the physics of the PSFs (which are defined in the phase space), we propose the following linear expansion scheme by translating the distortions in the phase domain to the principal components in the spatial domain.

Consider a simulator that generates a vector of random Zernike coefficients $\va = \{a_j\}_{j=1}^N$. Suppose the simulator has generated a set of $M$ vectors, $\{\va_1,\ldots,\va_M\}$. Since every $\va_i$ gives to a random phase, and every random phase corresponds to a random PSF, by generating $M$ Zernike coefficient vectors we will obtain a collection of $M$ random PSFs $\{\vh_1,\ldots,\vh_M\}$. By decomposing the random PSFs using the principal component analysis (PCA), we can extract a set of basis representations $\{\vu_j\}_{j=1}^p$ via
\begin{equation*}
    \{\vu_j\}_{j=1}^p = \mbox{PCA}\Big\{\vh_1,\ldots,\vh_M\Big\}.
\end{equation*}
Thus for every $\vh$, there exist $\{w_j\}_{j=1}^p$ such that
\begin{equation}
    \vh = \sum_{j=1}^p w_j \vu_j,
\end{equation}
where $w_j$ is the basis expansion coefficient. In other words, we converted the linear expansion of the phase numerically to a linear expansion of the PSF. This is a data-driven procedure, where the data comes from the physical model of the turbulence.

\textbf{Removing the tilts}. The above linear expansion model is valid when $\vy$ is the raw turbulent distorted image. However, since $\vy$ is the lucky frame, additional modeling of the other parts of the pipeline is needed. 

Referring back to the random phase concept in \eref{eq: Zernike}, we note that since the optical flow has compensated for the pixel movements, the tilts of the phase can be approximately dropped. 
More precisely, we consider a random phase $\phi(\vf)$ generated by a set of Zernike coefficients $\{a_1,\ldots,a_N\}$. Using \cite[Eq. 13]{Our_paper}, we define the tilt-compensated phase as
\begin{equation}
    \varphi(\vf) = \phi(\vf) - \valpha^T\vf,
\end{equation}
for some linear terms $\valpha$. The new PSFs defined through the tilt-compensated phase $\varphi(\vf)$ are center-aligned, which can be interpreted as outputs of the optical flow. When synthesizing the simulated training data, the tilts can be removed by setting $a_2 = a_3 = 0$. 

\begin{figure}[t]
	\centering
	\includegraphics[width=\columnwidth]{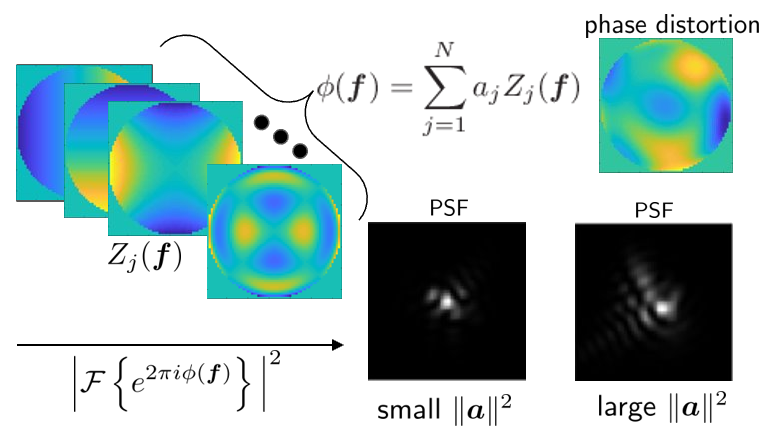}
	\vspace{-2ex}
	\caption{\textbf{Zernike representation of phase}. The phase $\phi(\vf)$ can be expressed as a linear combination of the Zernike basis $Z_j(\vf)$, with basis coefficients $a_j$. If the coefficient vector $\va = [a_j]$ has a large magnitude, then the distortion in the resulting PSF will be large.}
	\label{fig: psf_formation}
\end{figure}

After removing the tilts, the lucky fusion step is used to pick select patches according to two criteria we described in the lucky fusion section. To model this during the training data synthesis, we note that the ``wildness'' of the PSF is determined by the magnitude of the Zernike coefficients as shown in \fref{fig: psf_formation}. Therefore, to obtain good patches, we perform an importance sampling by rejecting large Zernike coefficients. Specifically, we define $\va_{4:N} = \{a_4,\ldots,a_N\}$ as the vector of Zernike coefficients from $a_4$ (because $a_2 = a_3 = 0$, and $a_1$ is a constant which does not affect the shape of the PSF.) We check whether $\|\va_{4:N}\|^2 \le \kappa$ for some threshold $\kappa$. If the magnitude $\|\va_{4:N}\|^2$ is above the threshold, we say that the PSF should not be used to form the PCA basis because lucky fusion will likely skip this PSF.

The proposed modeling of the optical flow and the lucky frame is justified by the optics and turbulence literature. In particular, tilt-compensation is a standard practice in adaptive optics where the tilt vector $\valpha$ is measured by phase offsets. When averaged over many random realizations, the tilt-compensated PSFs will give the short-exposure PSF. The small coefficient approach is also justified. It was first analyzed by Fried in \cite{Fried78}, where the threshold was 1 rad$^2$ for the Karhunen–Lo\`{e}ve expansion of the phase. In our case, the principle remains but the threshold is set empirically.

\textbf{Updating $\vh$}. The update of the PSF in \eref{eq: vv subproblem} now becomes
\begin{align}
\vw^{k+1} &= \argmin{\vw} \; \left\|\vy - \left(\sum_{j=1}^p w_j \vu_j\right) \circledast \vz^{k} \right\|^2 + \gamma r(\vw)  \label{eq: vw subproblem 1} \\
\vh^{k+1} &= \sum_{j=1}^p w_j^{k+1} \vu_j.
\label{eq: vw subproblem 2}
\end{align}
The first step \eref{eq: vw subproblem 1} defines the weight of the latent PSF. The prior for the weight is denoted as $r(\vw)$ for some regularization function $r(\cdot)$. The second step \eref{eq: vw subproblem 2} updates the PSF using the linear expansion concept. Depending on the prior $r(\cdot)$, e.g., $\ell_1$ prior, \eref{eq: vw subproblem 1} can be solved using standard optimization packages, e.g., CVX \cite{cvx} or the alternating direction methods of multiplier \cite{Boyd2011,chan2017plug}.

\textbf{Defining $r(\vw)$}. The prior distribution $r(\vw)$ can be determined empirically. Consider the set of generated PSFs $\{\vh_1,\ldots,\vh_M\}$. For each PSF $\vh_i$, we seek the \emph{sparest} representation using the bases $\{\vu_j\}_{j=1}^p$. We are interested in such representation because during the blind deconvolution step, the current estimate of the PSF is never perfect. Sparse representation entails the most influential bases while rejecting potential residue errors.

\begin{figure}[th]
	\centering
	\includegraphics[width=\linewidth]{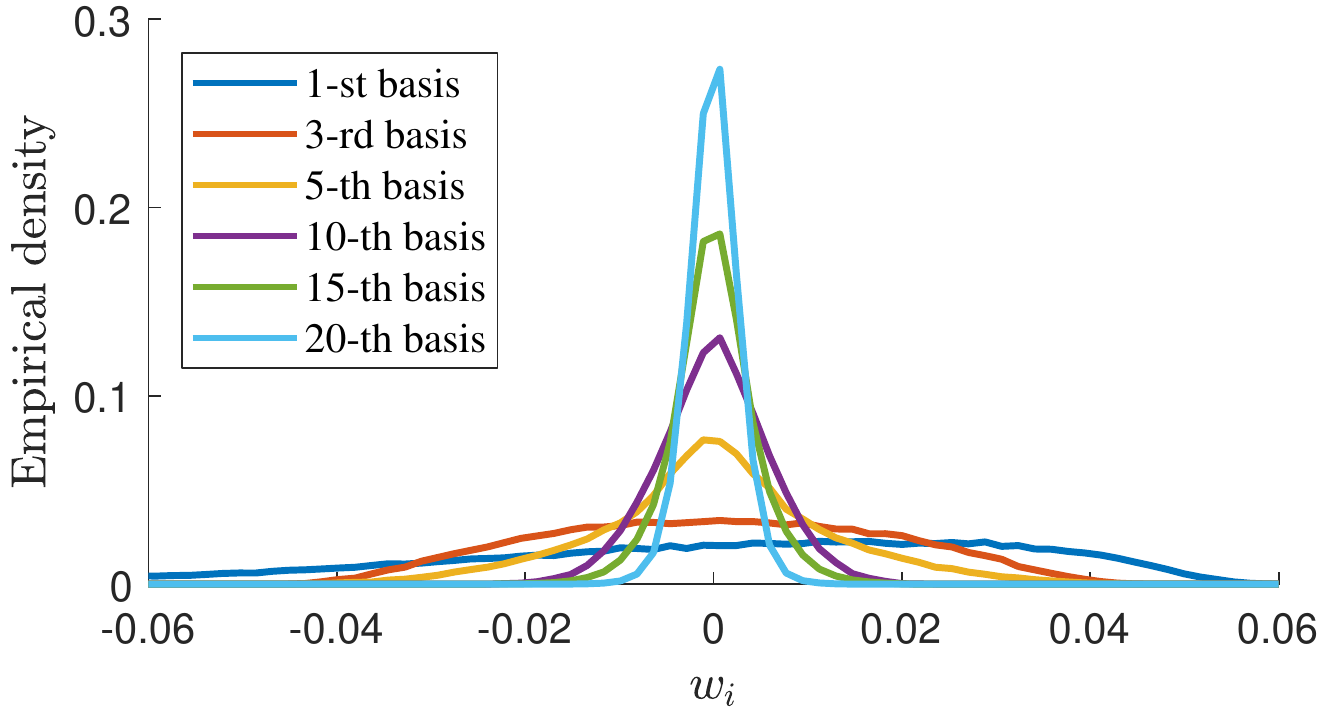}
	\caption{\textbf{Distribution of $\vw$}. Empirical distributions of the weights by solving \eref{eq: w optimizaiton} with 100,000 different simulated PSF $\vh$. The empirical distribution demonstrates a double-sided exponential distribution.}
	\label{fig:prior distribution}
\end{figure}

The regularization function is defined as
\begin{equation}
r(\vw) = \sum_{j=1}^p \frac{|w_j|}{\sigma_j},
\label{eq: r(w)}
\end{equation}
where $\sigma_i$ is the standard deviation of $w_i$ learned from data. This is equivalent to assuming that the prior distribution of $\vw$ follows a field-of-expert model:
\begin{equation}
    p(\vw) = \exp\left\{-\sum_{j=1}^p \frac{|w_j|}{\sigma_j}\right\},
\end{equation}
such that $r(\vw) = -\log p(\vw)$. Note that the regularization in Equation \eref{eq: r(w)} is a weight $\ell_1$ norm. The $\ell_1$ norm is used here as a compromise between model fidelity and complexity. Norms other than $\ell_1$ can potentially lead to better fitting (as in the mean squared fitting error), but the computational complexity of solving the corresponding optimization will become higher. The choice of the $\ell_1$ norm is further justified by the shape of the empirical distributions shown in \fref{fig:prior distribution}. We observe that the distributions indeed follow an exponentially decaying function, and has a tail heavier than a Gaussian.

To find the sparse representation, we solve the following problem offline (during the training phase):
\begin{equation}
\vw^{(i)} = \argmin{\vw}\;\; \|\vw\|_0 \;\; \mbox{s.t.} \;\; \left\|\vh_i - \sum_{j=1}^p w_j \vu_j \right\|^2 \le \tau,
\label{eq: w optimizaiton}
\end{equation}
where $\tau$ is a pre-define threshold. Note that Equation \eref{eq: w optimizaiton} is a standard $\ell_0$ basis pursuit denoise problem, which can be solved using tools such as orthogonal matching pursuit \cite{Pati1993}. After solving the problem, we will have a collection of the expansion coefficients $\{\vw^{(i)}\}_{i=1}^M$. Plotting the empirical density of the collection $\{w_j^{(1)},\ldots,w_j^{(M)}\}$ will give us plots shown in \eref{fig:prior distribution}, for each of the $j$-th basis. The empirical distributions will then define the regularization function $r(\vw)$.

\begin{table}[t]
\caption{Runtime (sec) for generating 100 images of sizes $256\times 256$}
\label{table: runtime}
\centering
\begin{tabular}{l r}
\hline\hline
Reference Frame &    3.17 sec\\
Optical Flow    & 260.02 sec\\
Lucky Fusion    &   6.71 sec\\
Blind Deconvolution \hspace{4ex}  & 336.44 sec\\
\hline
Overall         & 606.34 sec\\
\hline
\end{tabular}
\end{table}

\subsection{Computing Time} 

We report the computing time required to process a sequence with 100 $256\times256$ frames in Table \ref{table: runtime}. The optical flow \cite{Liu2009optflow} is implemented in C++ and other steps are implemented in MATLAB. The algorithm is tested on an Intel I7-7700HQ CPU with 16 GB memory. It is observed that the bottle-neck of the run time is the optical flow, where 100 frame-to-frame image alignment needs to be done. The other bottleneck is the blind deconvolution, where the alternating minimization is used. Compared to existing methods which also use optical and blind deconvolution, the additional cost of the proposed method is the reference generation step. However, this step contributes to less than 1\% of the total runtime.

\section{Experimental Results}
\label{sec:experiments}

\subsection{Evaluation Data}
We evaluate the proposed method on three types of data:
\begin{itemize}
    \item \textbf{Synthetic sequences}, where we have ground truths to compute the PSNRs. Our simulation is based on the method presented in \cite{Our_paper}, with simulation parameters listed in Table~\ref{table: parameters}. The refractive index constant $C_n^2$ ranges from $1\times 10^{-15}$m$^{-2/3}$ to $5 \times 10^{-15}$m$^{-2/3}$, which correspondingly makes the aperture-to-coherence ratio $D/r_0$ ranges between 1 and 4. 
    \item \textbf{Real Hot-air Sequences}, popularly used in the computer vision literature, e.g., Hirsch et al. \cite{Hirsch2010} and Anantrasirichai et al. \cite{Anantrasirichai2013}. These sequences are collected by placing the camera behind a hot-air burner. These hot-air sequences are good surrogates of the true long-range turbulence, but there are limitations. 
    \item \textbf{Real Long-range Sequences}, collected on 25 September, 2019. The path length between the camera and the object is approximately 4km at a temperature of 30$^\circ$C. In addition, we also use the NATO RTO SET/RTG-40 dataset reported by Leonard, Howe and Oxford in \cite{Leonard_Howe_Oxford}. 
\end{itemize}

All the sequences used in this paper contain 100 frames, which are consistent with \cite{Anantrasirichai2013,Milanfar2013,Lou2013,Hirsch2010}.

\begin{table}[h]
\caption{Simulation Parameters}
\label{table: parameters}
\centering
\begin{tabular}{ll}
\hline\hline
Parameter & Value \\
\hline
Path length         & $L$ = 4km\\
Aperture Diameter   & $D$ = 0.1m\\
Focal Length        & $d$ = 0.4m\\
Wavelength          & $\lambda$ = 525nm\\
Zernike Phase Size  & $64 \times 64$ pixels\\
Image Size          & $512 \times 512$ pixels\\
Nyquist spacing (object plane) $\frac{L\lambda}{2D}$ & $\delta_0 = 1.05$mm\\
Nyquist spacing (focal plane) $\frac{d\lambda}{2D}$ & $\delta_f = 1.05\mu$m\\
\hline
\end{tabular}
\end{table}

\begin{figure}[t]
	\centering
	\begin{tabular}{ccc}
		\includegraphics[width=0.32\linewidth]{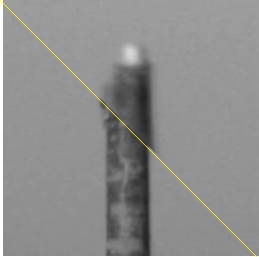}&
		\hspace{-2ex}\includegraphics[width=0.32\linewidth]{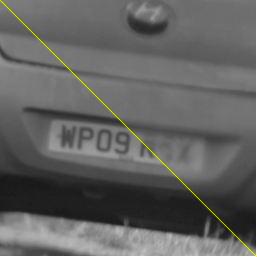}&
		\hspace{-2ex}\includegraphics[width=0.32\linewidth]{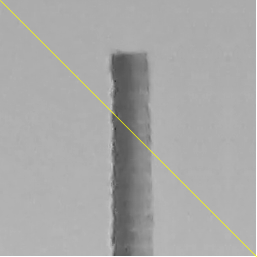}\\
		(a) Hot-air & (b) Hot-air  & (c) Long-range
	\end{tabular}
	\caption{\textbf{Comparing hot-air and real turbulence}. Hot-air sequences can have really ``lucky'' frames, but a real turbulence sequence is consistently degraded. Thus, just using hot-air sequences for evaluation is inappropriate. For each sub-figure, the upper right is a typical frame randomly picked from a sequence, and the bottom left is a manually selected lucky frame from the same sequence. 	
	(a) \texttt{Chimney} \cite{Hirsch2010} (hot-air), (b) \texttt{A3. Back car} \cite{Anantrasirichai2013} (hot-air), and (c) \texttt{New Chimney} (long-range). }
	\label{fig:hotair}
\end{figure}

We like to comment on the difference between a hot-air sequence and the long-range sequences. \fref{fig:hotair} shows three video sequences. The first two sequences \texttt{Chimney} \cite{Hirsch2010} and \texttt{A3. Back car} \cite{Anantrasirichai2013} are created by hot-air burners, and the third sequence \texttt{New Chimney} is a real long-range turbulence captured at 4km. In each of the sub-figures, we show a typical frame and a manually selected ``lucky'' frame. For the two hot-air sequences, we observe that the ``lucky'' frames actually have very high quality, so high-quality that even post-processing becomes unnecessary. The reason for such phenomenon is that the turbulence caused by a hot-air burner is isoplanatic which is quite different from the real long-range turbulence. If we look at the real long-range turbulence case in \fref{fig:hotair}, we observe that the best lucky frame is still quite similar to the other frames.

\begin{figure}[t]
	\centering
	\includegraphics[width=\linewidth]{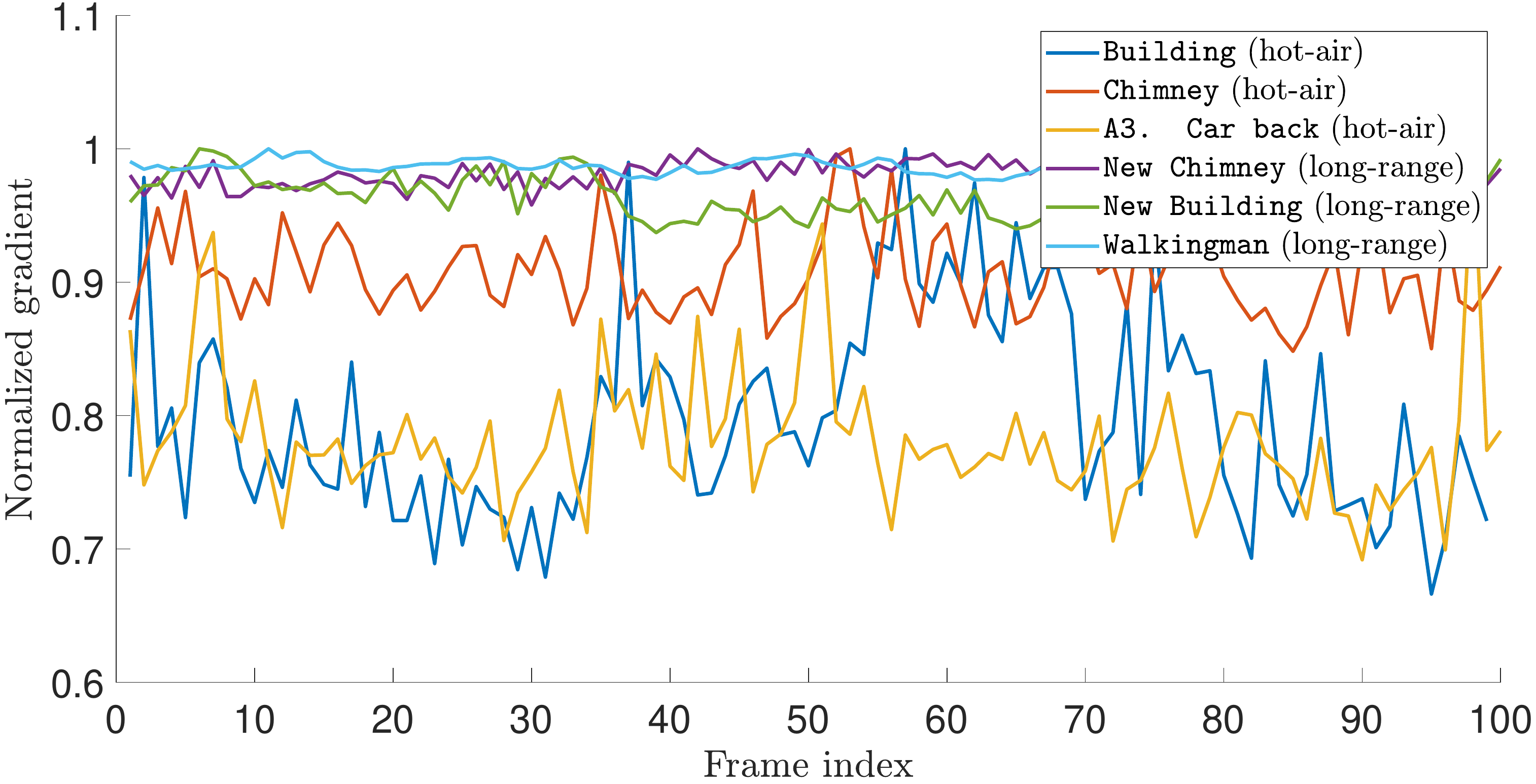}
	\caption{\textbf{Normalied gradient for hot-air and long range sequences}. Short range hot-air sequences exhibit a strong fluctuation in the the temporal normalized gradients, whereas long range turbulence sequences do not.}
	\label{fig:sharpness}
\end{figure}

The difference between the hot-air and the long-range turbulence can be quantitatively compared. Consider the \emph{normalized gradient}
\begin{align}
    \calE_t \bydef \frac{\sum_{\vr}\|\nabla \vy(\vr,t)\|_1}{\max\limits_{t} \sum_{\vr} \|\nabla \vy(\vr, t)\|_1}, 
\end{align}
which is a metric measuring the sharpness of the frames (relative to the sharpest frame in the sequence). If we plot $\calE_t$ as a function of time, we observe that long-range turbulence has substantially different behavior than hot-air. In particular, the curves in \fref{fig:sharpness} show that ``lucky frames'' generally do not appear often in long-range sequences, but happens occasionally in hot-air sequences. As a result, we argue that it is necessary to evaluate a method using not just the hot-air sequences but also the real long-range turbulence sequences.

\subsection{Evaluating Reference Frames}
We evaluate the effectiveness of the proposed reference frame extraction method using two sets of experiments.

\begin{figure}[th]
	\centering
	\begin{tabular}{c c c c c c}
    	\includegraphics[width=0.15\linewidth]{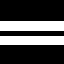} & 
		\hspace{-2ex}\includegraphics[width=0.15\linewidth]{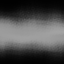} &
		\hspace{-2ex}\includegraphics[width=0.15\linewidth]{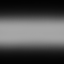} &
		\hspace{-2ex}\includegraphics[width=0.15\linewidth]{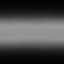} &
		\hspace{-2ex}\includegraphics[width=0.15\linewidth]{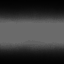} &
		\hspace{-2ex}\includegraphics[width=0.15\linewidth]{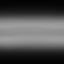} \\
        \footnotesize{(a) Truth} & \hspace{-2ex} \footnotesize{(b) Input} & \hspace{-2ex} \footnotesize{(c) Avg} & \hspace{-2ex} \footnotesize{(d) \cite{Lin2013}} & \hspace{-2ex} \footnotesize{(e) \cite{Lau2017}} & \hspace{-2ex} \footnotesize{(f) Ours} 
	\end{tabular}
	
	\begin{tabular}{c c}
		\includegraphics[width=0.45\linewidth]{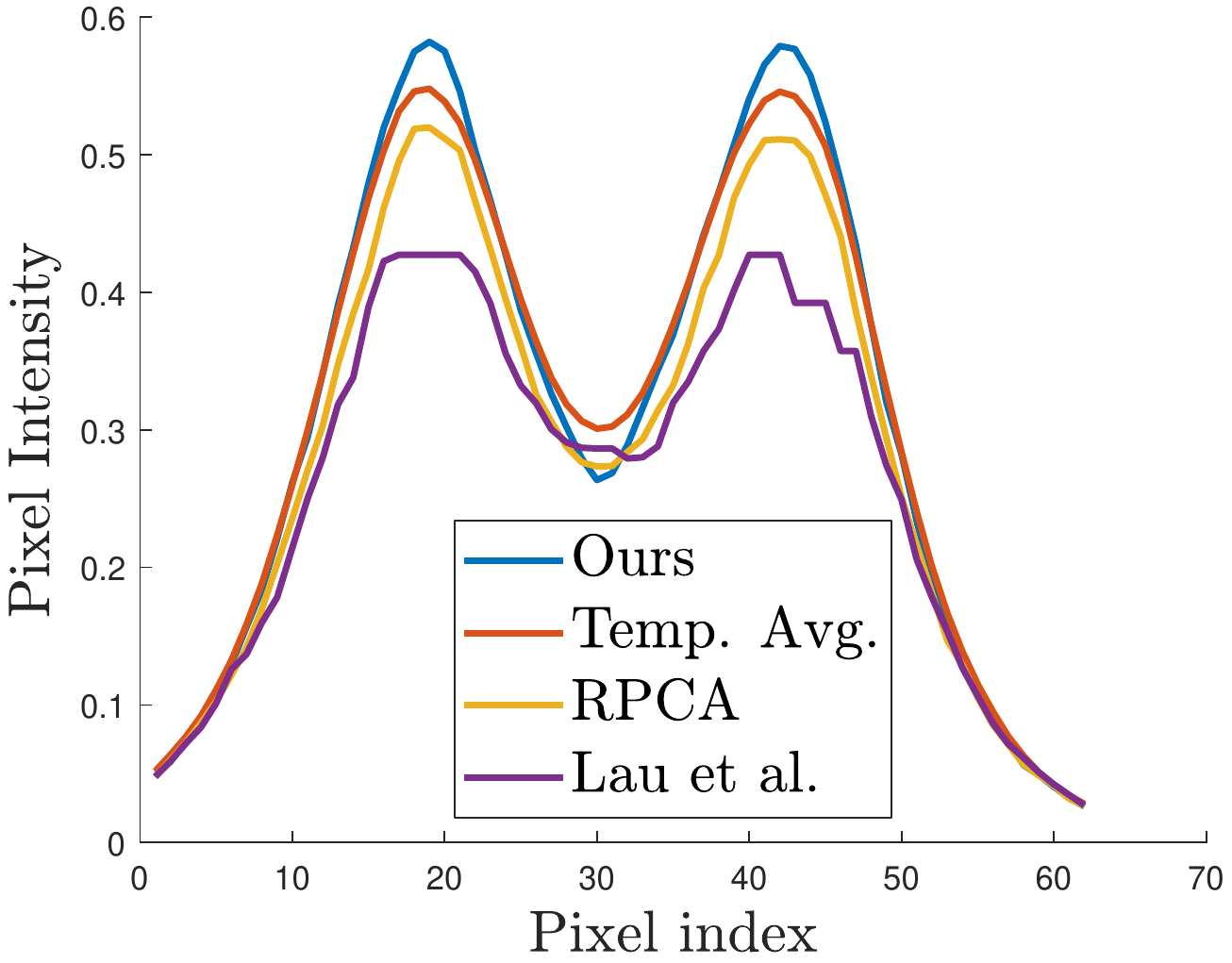} &
		\includegraphics[width=0.45\linewidth]{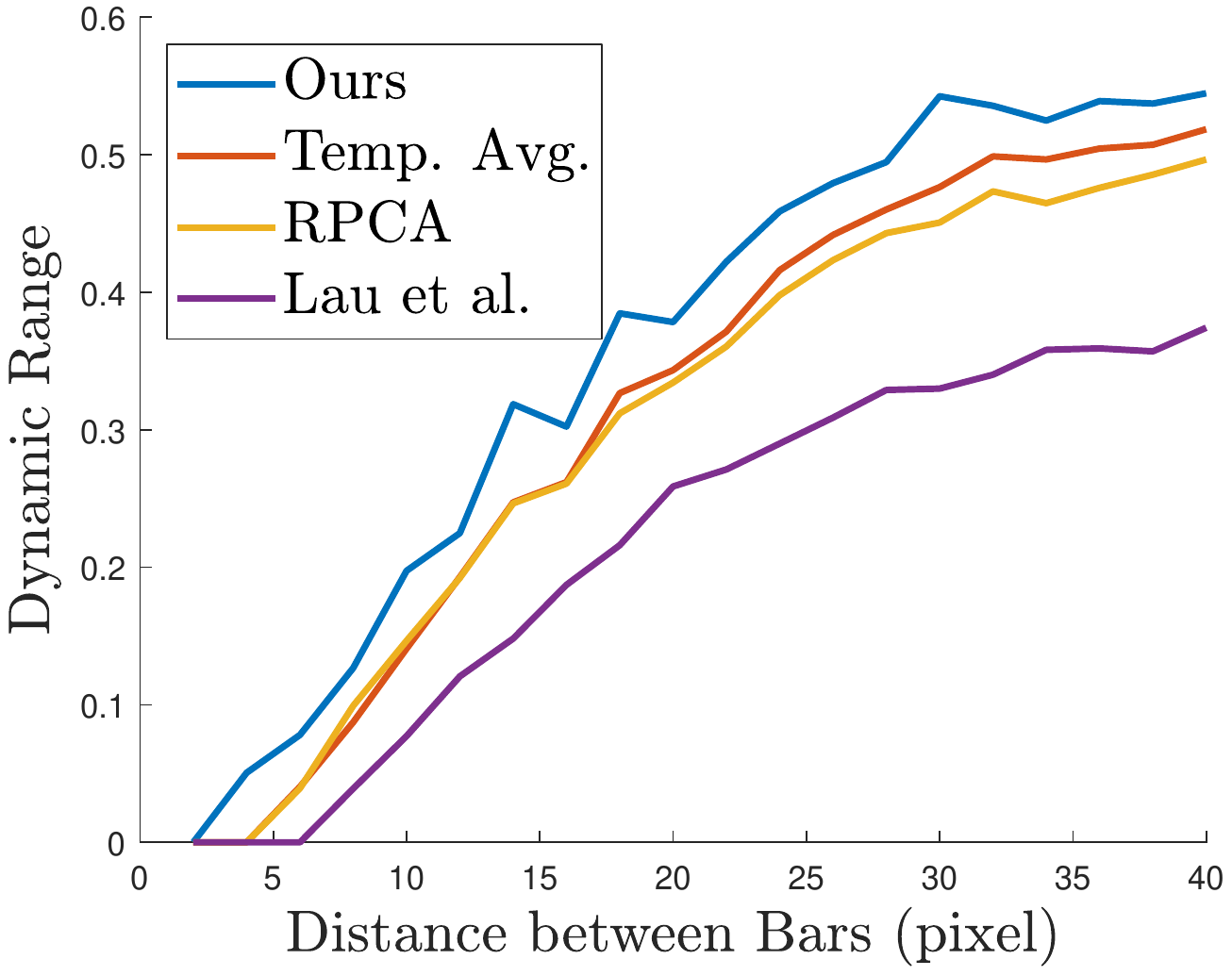} \\
        \footnotesize{(g) Intensity Plot} & \footnotesize{(h) Dynamic Range}
	\end{tabular}
	\caption{\textbf{Evaluating the reference frame method}. The objective of this experiment is to evaluate which reference frame reconstruction method can achieve better resolution. (a) Ground truth. Spacing between bars is 4 pixel. (b) A simulated frame (c) Temporal average. (d) RPCA (e) Lau's reference frame (f) Our reference frame (g) Dynamic range of the reference frame with varying spacing (h) The cross-section at center of the bars when the spacing is 10. }
	\label{fig:ref_drange}
\end{figure}

\textbf{Resolution enhancement}. The first experiment is a synthetic experiment that tests the minimal resolvable distances. We consider a square image consisting of two horizontal lines separated by a distance of 10 pixels, as shown in \fref{fig:ref_drange}(a). We apply the simulator presented in \cite{Our_paper}, with parameters specified in Table \ref{table: parameters} and $C_n^2 = 1.5\times10^{-15}$m$^{-2/3}$. \fref{fig:ref_drange}(b) shows one random realization of the turbulent distorted image. The reference frames generated by competing methods are shown in \fref{fig:ref_drange}(c) - \fref{fig:ref_drange}(f). For the RPCA method by Lin et al. \cite{Lin2013}, default parameters are used. For the methods by Lau et al. \cite{Lau2017}, we retain 25\% of the original frames as suggested by the authors. For our proposed method, we use a $7 \times 7$ space-time search window. As can be seen in the figures, the proposed method has the best recovery.

\begin{figure}[h]
	\centering
	\begin{tabular}{cc}
	 \hspace{-2ex}
	    \includegraphics[width=0.45\linewidth]{pix/ref/zoomed_input.png}&
        \hspace{-2ex} \includegraphics[width=0.45\linewidth]{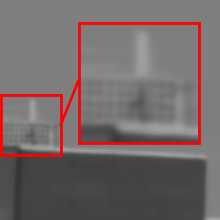} \\
        (a) Input & (b) Lin et al. \cite{Lin2013}\\
        \hspace{-2ex} \includegraphics[width=0.45\linewidth]{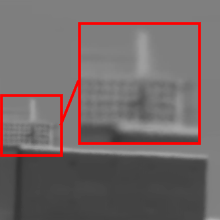} &
        \includegraphics[width=0.45\linewidth]{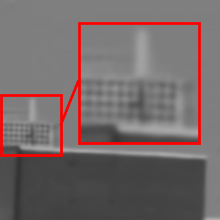}\\
        \hspace{-2ex} (c) Lau et al. \cite{Lau2017}&
        \hspace{-2ex} (d) Proposed
	\end{tabular}
	\caption{\textbf{Evaluating the reference frame method using real static data}. This figure shows the reconstructed references frames by competing methods.}
	\label{fig:ref_visual}
\end{figure}

\begin{figure*}[h]
	\centering
	\begin{tabular}{ccc}
		\hspace{-2ex}
		\includegraphics[trim={3cm 0 2cm 0},clip, width=0.3\linewidth]{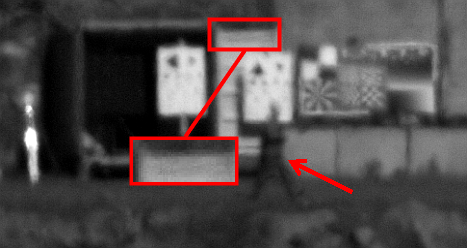} &
		\hspace{-2ex}
		\includegraphics[trim={3cm 0 2cm 0},clip, width=0.3\linewidth]{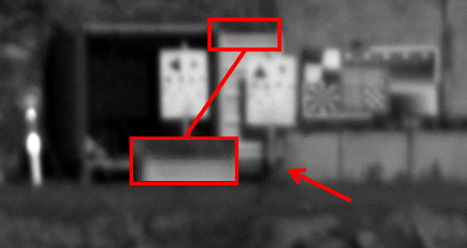} &
		\hspace{-2ex}
		\includegraphics[trim={3cm 0 2cm 0},clip, width=0.3\linewidth]{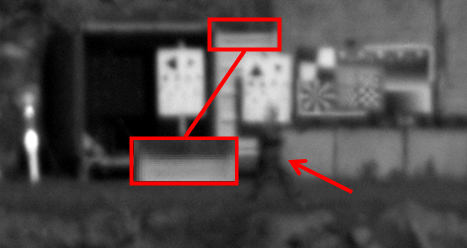} \\
		\hspace{-2ex} \small{(a) Raw input} &
		\hspace{-2ex} \small{(b) Temp. Avg} &
		\hspace{-2ex} \small{(c) Proposed}
	\end{tabular}
	\caption{\textbf{Evaluating the reference frame method using real dynamic data}. (a) Raw input. (b) Temporal averaging. Observe that the man is blurred over 100 frames. (c) Ours. Observe that the man remains in the image while the background is stabilized.}
	\label{fig:ref_frame_comp 2}
\end{figure*}

 \begin{figure*}[!]
	\centering
	\begin{tabular}{cccccc}
		\hspace{-2ex}\includegraphics[width=0.15\linewidth]{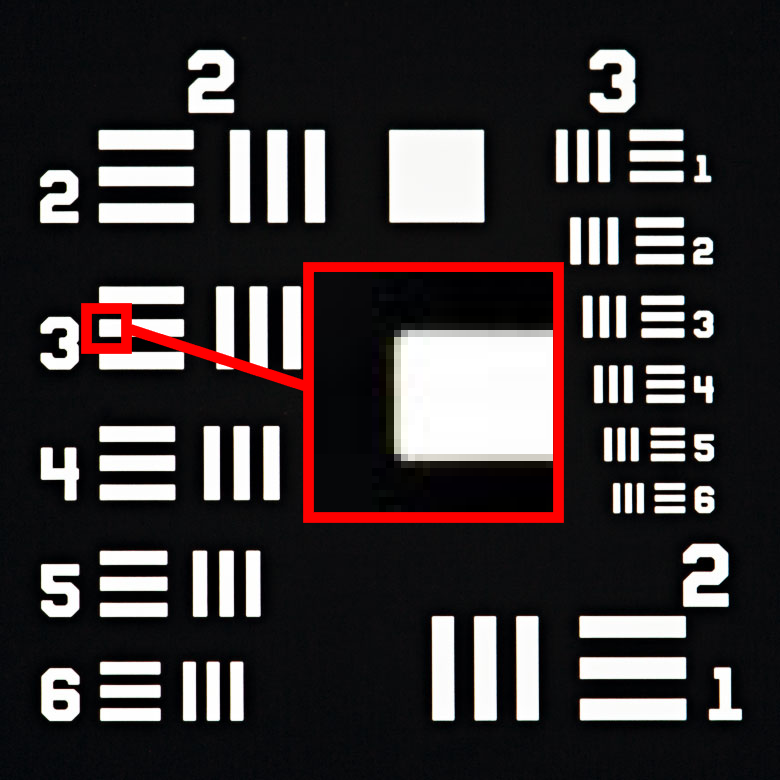}   & \hspace{-2ex} \includegraphics[width=0.15\linewidth]{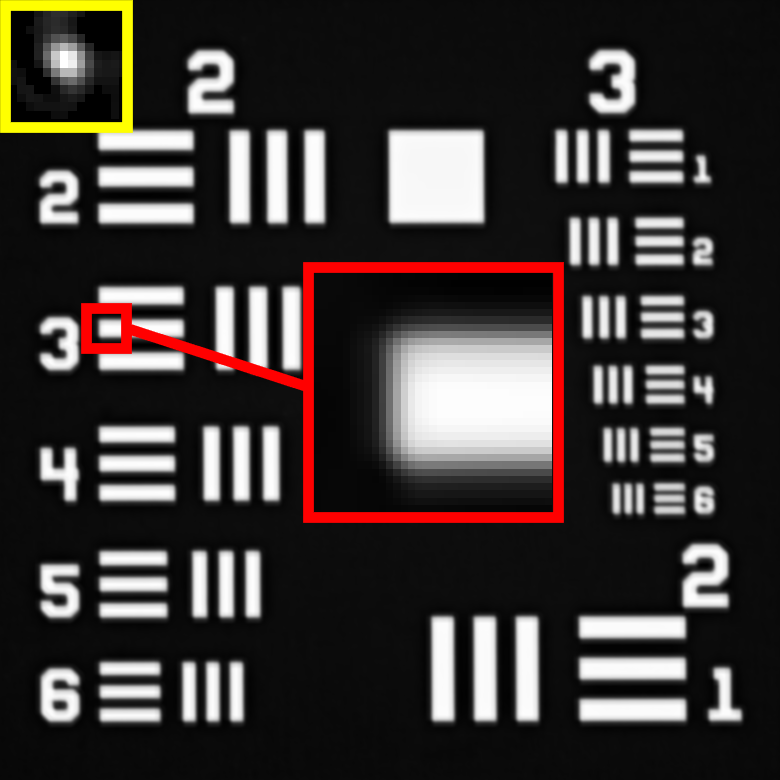} & \hspace{-2ex}
		\includegraphics[width=0.15\linewidth]{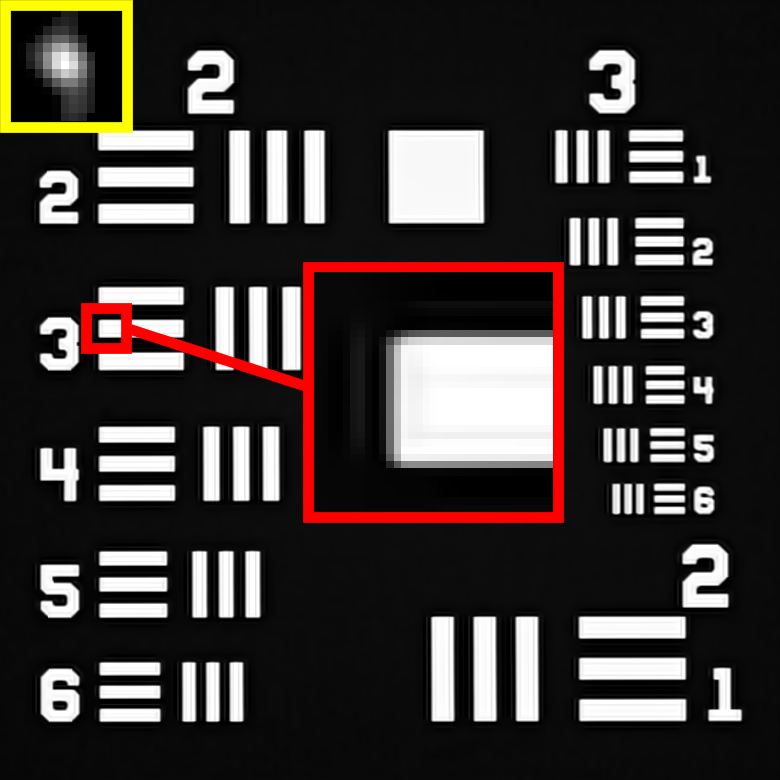} & \hspace{-2ex}
		\includegraphics[width=0.15\linewidth]{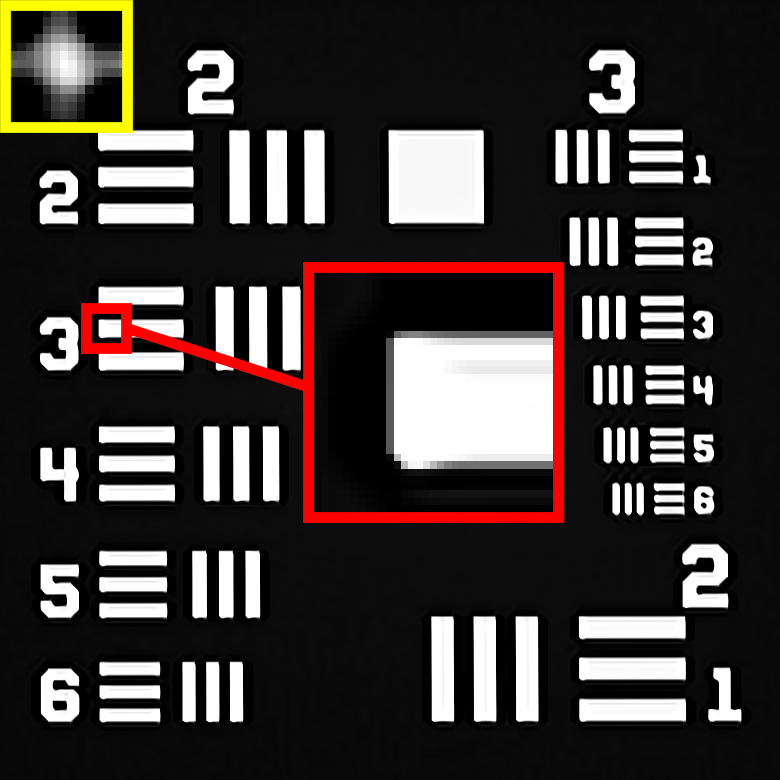} & \hspace{-2ex}
		\includegraphics[width=0.15\linewidth]{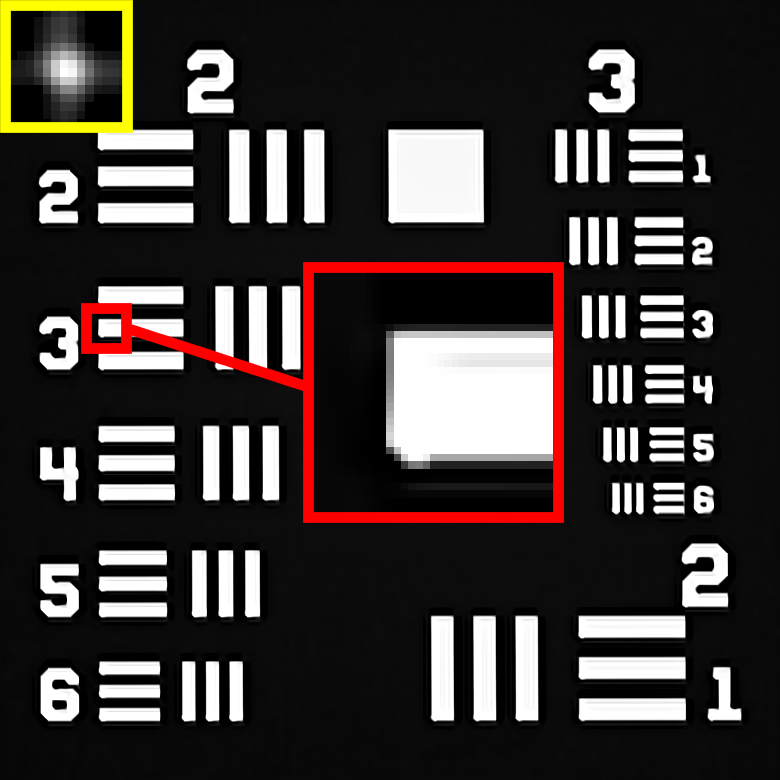} & 
		\hspace{-2ex}
		\includegraphics[width=0.15\linewidth]{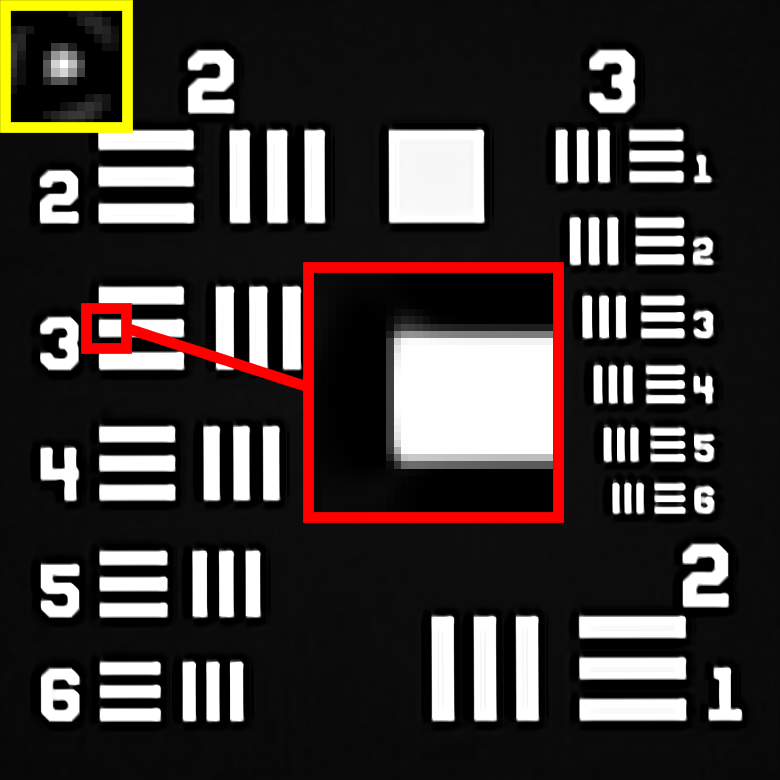}  \\
		\hspace{-2ex}\small{(a) Ground Truth} & 
		\hspace{-2ex} \small{(b) Blur Input} &
		\hspace{-2ex} \small{(c) \cite{Shan2008}: 23.34 dB} &
		\hspace{-2ex} \small{(d) \cite{Chakrabarti2016}: 22.99 dB} & 
		\hspace{-2ex} \small{(e) \cite{Xu2018}: 23.75 dB} &
		\hspace{-2ex} \small{(f) Proposed: 24.78 dB} \\
	\end{tabular}
	\caption{\textbf{Comparing blind deconvolution using synthetic data}. In this example we use the USAF resolution chart. We blur the ground truth using an ideal short-exposure PSF, and reconstruct the images using different methods. }
	\label{fig:USAFresult}
\end{figure*}

A more quantitative result can be seen in \fref{fig:ref_drange}(g), where we plot the cross-section of the estimated reference frames. In \fref{fig:ref_drange}(h) we further plot the dynamic range of the recovered pattern, which is defined as the gap between the peak and the valley of the intensity plots. As we can see in the intensity plot and the dynamic range plot, the proposed method generates the best contrast of the patterns. The RPCA by Lin et al. \cite{Lin2013} and the frame selection method by Lau et al. \cite{Lau2017} both adopt a low rank approximation approach. As such, the recovered patterns have lower intensity values because the sparse components are subtracted from the images.

\textbf{Test on real data}. In \fref{fig:ref_visual} we show a visual comparison of the extracted reference frames from a real turbulence sequence, where we compare the RPCA by Lin et al. \cite{Lin2013}, the frame selection by Lau et al. \cite{Lau2017}, and the proposed method. As we can see in the results, the proposed method generates a noticeably sharper reference frame.

In \fref{fig:ref_frame_comp 2} we show a sequence containing a moving object. Because of the moving object, RPCA essentially becomes the temporal average when we consider the largest principal component. Our result in \fref{fig:ref_frame_comp 2} shows that the temporal average will wash out the moving object. In contrast, the proposed method can retain it.

\subsection{Evaluating Blind Deconvolution}
We decouple the blind deconvolution from the rest of the pipeline and evaluate the effectiveness of the proposed basis expansion idea. In this experiment, we compare with a classic method by Shan et al. \cite{Shan2008} and two recent deep nerual networks by Chakrabarti \cite{Chakrabarti2016} and Xu et al. \cite{Xu2018}. The codes and models were obtained from the authors' websites. We use pre-trained models for the two neural networks, and fine-tune the parameters in \cite{Shan2008}. For our proposed method, we fixed $\lambda = 0.05$ and $\gamma = 1\times10^{-3}$ for all experiments reported in this section. 

In \fref{fig:USAFresult} we show a visual comparison using the USAF resolution chart with $D/r_0 = 1.4$. We blur the ground truth image using an ideal short-exposure PSF, and apply various blind deconvolution methods to recover the blur and the image. As we can see, the proposed method generates an image containing the least amount of artifacts. The estimated PSF is also more structured and interpretable than the neural network methods.

\begin{figure}[h]
	\centering
	\includegraphics[width=0.9\linewidth]{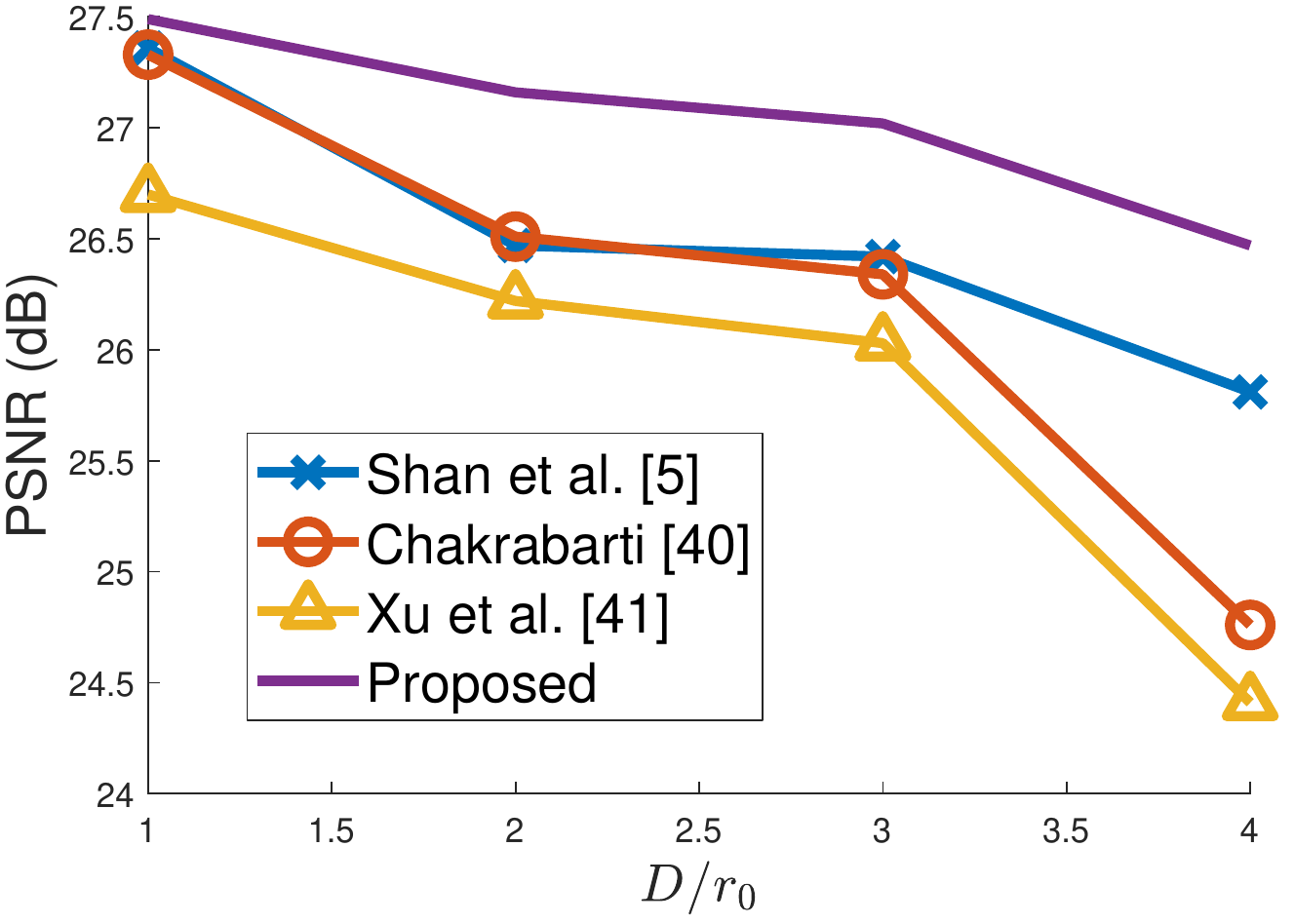}
	\caption{\textbf{PSNR curves of blind deconvolution using synthetic data}. The PSNRs are averaged over 24 images of the Kodak image dataset, and every image is tested 50 times for different random turbulence PSFs. Thus every data point reported in this table is an average over 1200 testing scenarios. }
	\label{fig:blind_deconv}
\end{figure}

We evaluate the result using the 24 images in the Kodak image dataset. Each image is blurred with 50 random point spread functions, where each one is constructed by averaging 10 short-exposure PSFs under different turbulence levels (with $D/r_0$ from 1 to 2.6 with step size 0.4). This creates a total of 1200 testing images. The average PSNR value with respect to the ground truth under different turbulence levels is reported in \fref{fig:blind_deconv}. It is evident from the table that the proposed method has considerably better reconstruction PSNR than the competing methods. We argue that this is due to the improved physics-inspired prior for the point spread functions.

\subsection{Evaluating the Overall Pipeline}
We now evaluate the effectiveness of the overall proposed algorithm by comparing it with several existing methods: Sobolev gradient flow (SG) by Lou et al. \cite{Lou2013}, near-diffraction-limit (NDL) by Zhu and Milanfar \cite{Milanfar2013} with deblurring using Shan et al. \cite{Shan2008}, and wavelet fusion method (CLEAR) by Anantrasirichai et al. \cite{Anantrasirichai2013}. The codes are provided by the original authors and the internal parameters are tuned according to the best of our knowledge.

\begin{figure}[h]
	\centering
	\includegraphics[width=0.8\linewidth]{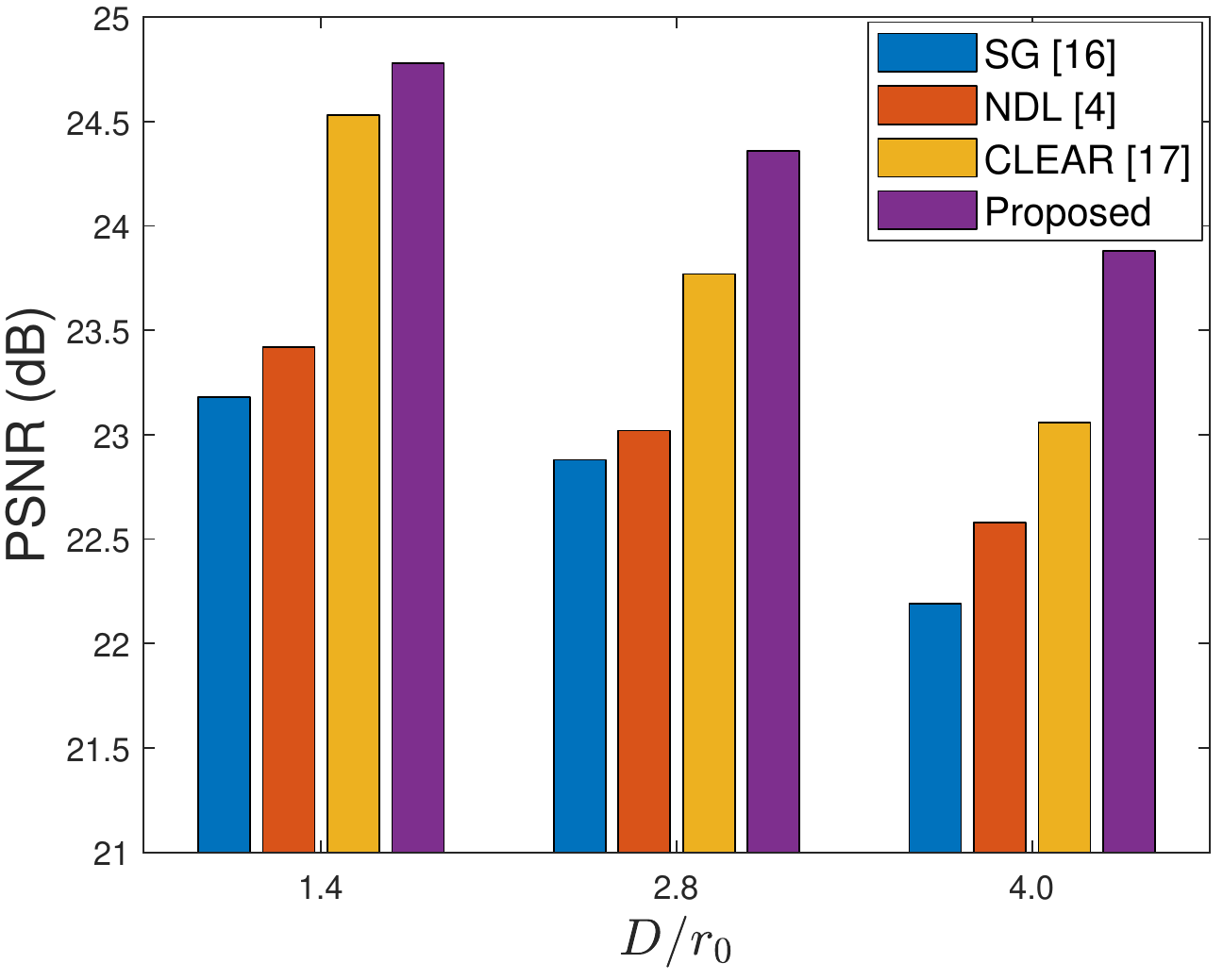}
	\caption{\textbf{PSNR of the entire pipeline using synthetic data}. The PSNRs reported here are averaged over 10 sequences. }
	\label{fig:synthetic}
\end{figure}

\begin{figure*}[t]
	\centering
	\begin{tabular}{c c c c c c}
	    \includegraphics[width=0.15\linewidth]{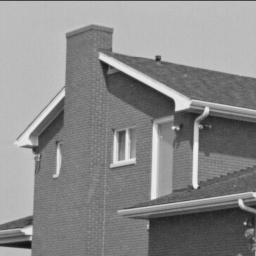}&
	   \hspace{-2ex}\includegraphics[width=0.15\linewidth]{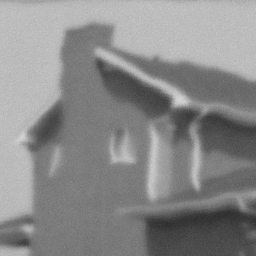} & \hspace{-2ex}\includegraphics[width=0.15\linewidth]{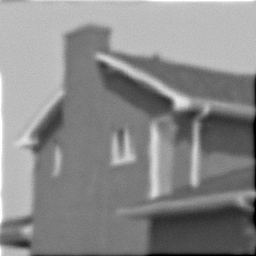} & \hspace{-2ex}\includegraphics[width=0.15\linewidth]{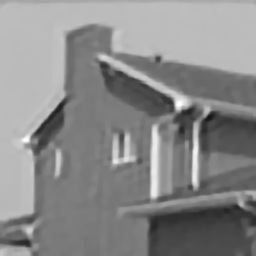} & \hspace{-2ex}\includegraphics[width=0.15\linewidth]{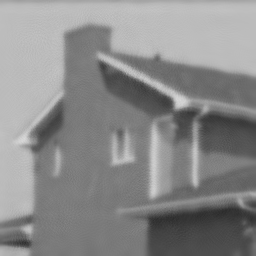} & \hspace{-2ex}\includegraphics[width=0.15\linewidth]{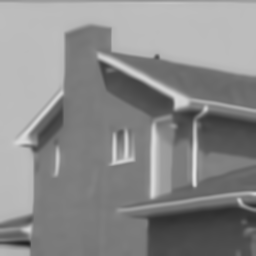} \\
       (a) Clean & \hspace{-3.5ex} (b) Observed & \hspace{-3.5ex} (c) SG \cite{Lou2013} & \hspace{-2ex} (d) NDL \cite{Milanfar2013} & \hspace{-2ex} (e) CLEAR \cite{Anantrasirichai2013} & \hspace{-2ex} (f) Proposed\\
       & 22.08dB & 22.36dB & 24.58dB & 23.93dB & 24.72dB
       \end{tabular}
       \vspace{-2ex}
	\caption{\textbf{Overall comparison using synthetic static data}. The sequence is distorted by a turbulence level of $D/r_0 = 2.8$. PSNR comparisons of the entire dataset is shown in \fref{fig:synthetic}.}
	\label{fig:houseResult}
\end{figure*}

\textbf{Synthetic dataset}. We synthetically generate turbulent sequences for 10 ``standard'' images (e.g., Lena, House, Cameraman, etc), where each sequence consists of 100 frames. The turbulence levels are set as $D/r_0 = $ 1.4, 2.8 and 4.0 respectively. The results are reported in \fref{fig:synthetic}, where each PSNR value corresponds to an average over the 10 testing sequences. A visual comparison of the House sequence is shown in \fref{fig:houseResult}. While all the algorithms yield similar results under a low turbulence level, our approach offers superior performance as the turbulence level increases.

\begin{figure*}[t]
    \centering
    \begin{tabular}{c c c c c c}
    	\hspace{\nspacetwo ex}\includegraphics[width=\nwidth\linewidth]{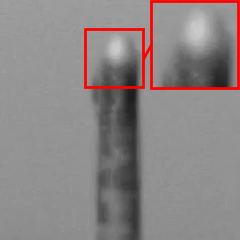} &
	   \hspace{\nspace ex}\includegraphics[width=\nwidth\linewidth]{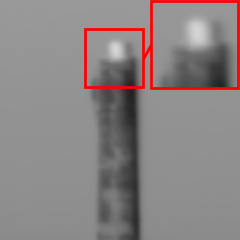} & 
	   \hspace{\nspace ex}\includegraphics[width=\nwidth\linewidth]{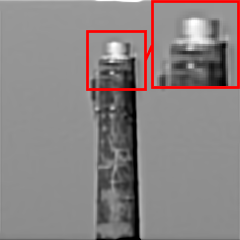} & 
	   \hspace{\nspace ex}\includegraphics[width=\nwidth\linewidth]{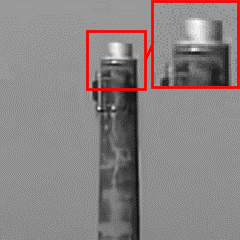} & 
	   \hspace{\nspace ex}\includegraphics[width=\nwidth\linewidth]{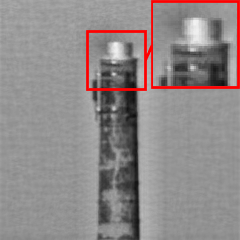} & 
	   \hspace{\nspace ex}\includegraphics[width=\nwidth\linewidth]{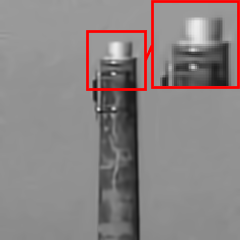} \\
	   \hspace{\nspacetwo ex}\includegraphics[width=\nwidth\linewidth]{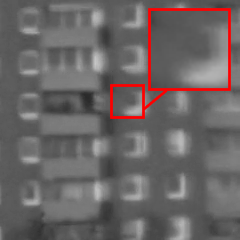} &
	   \hspace{\nspace ex}\includegraphics[width=\nwidth\linewidth]{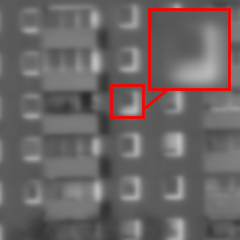} & 
	   \hspace{\nspace ex}\includegraphics[width=\nwidth\linewidth]{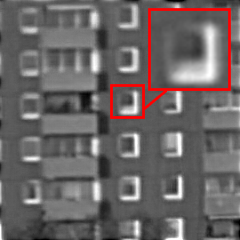} & 
	   \hspace{\nspace ex}\includegraphics[width=\nwidth\linewidth]{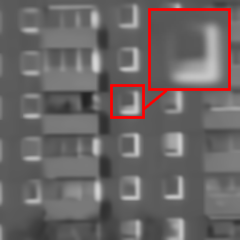} & 
	   \hspace{\nspace ex}\includegraphics[width=\nwidth\linewidth]{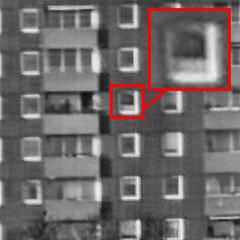} & 
	   \hspace{\nspace ex}\includegraphics[width=\nwidth\linewidth]{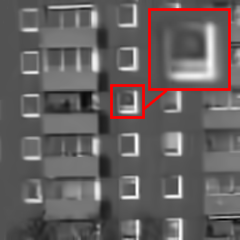} \\
	   (a) Observed & \hspace{-2ex} (b) Temp. Avg. & \hspace{-2ex} (c) SG \cite{Lou2013} & \hspace{-2ex} (d) NDL \cite{Milanfar2013} & \hspace{-2ex} (e) CLEAR \cite{Anantrasirichai2013} & \hspace{-2ex} (f) Proposed
    \end{tabular}
    \vspace{-1ex}
    \caption{\textbf{Overall comparison using real ``hot-air'' static sequences}. The sequences were previously reported in \cite{Hirsch2010}. [Top] The \texttt{chimney} . [Bottom] \texttt{Building}. Note that these are real videos where ground truths are not available.}
    \label{fig:realexp1}
\end{figure*}

\begin{figure*}[t]
    \centering
    \begin{tabular}{c c c c c c}
       \hspace{\nspacetwo ex}\includegraphics[width=\nwidth\linewidth]{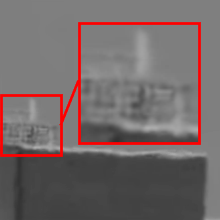} &
	   \hspace{\nspace ex}\includegraphics[width=\nwidth\linewidth]{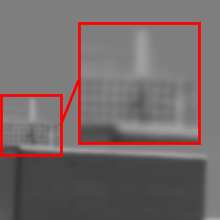} & \hspace{\nspace ex}\includegraphics[width=\nwidth\linewidth]{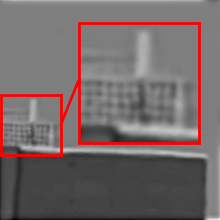} & \hspace{\nspace ex}\includegraphics[width=\nwidth\linewidth]{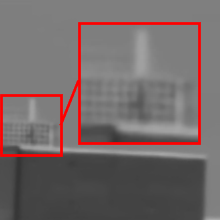} & \hspace{\nspace ex}\includegraphics[width=\nwidth\linewidth]{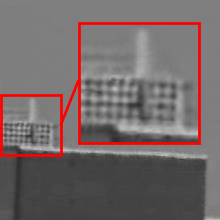} & \hspace{\nspace ex}\includegraphics[width=\nwidth\linewidth]{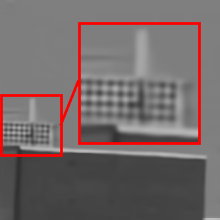} \\
	   \hspace{\nspacetwo ex}\includegraphics[width=\nwidth\linewidth]{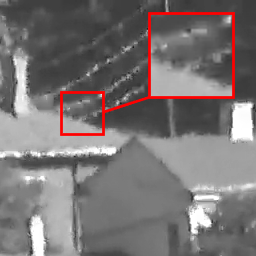} &
	   \hspace{\nspace ex}\includegraphics[width=\nwidth\linewidth]{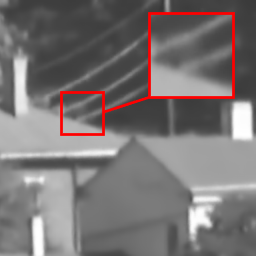} & \hspace{\nspace ex}\includegraphics[width=\nwidth\linewidth]{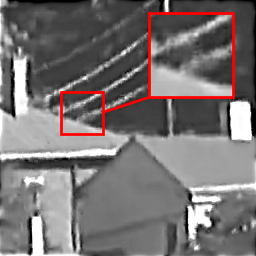} & \hspace{\nspace ex}\includegraphics[width=\nwidth\linewidth]{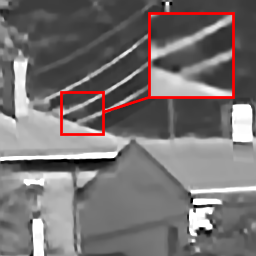} & \hspace{\nspace ex}\includegraphics[width=\nwidth\linewidth]{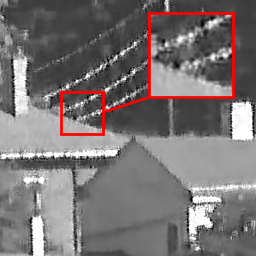} & \hspace{\nspace ex}\includegraphics[width=\nwidth\linewidth]{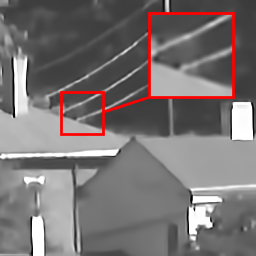} \\
	   \hspace{\nspacetwo ex}\includegraphics[width=\nwidth\linewidth]{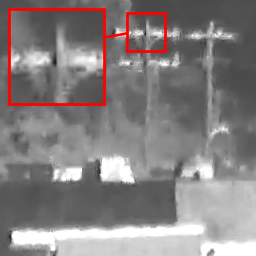} &
	   \hspace{\nspace ex}\includegraphics[width=\nwidth\linewidth]{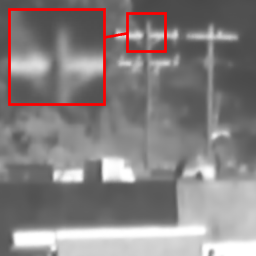} & \hspace{\nspace ex}\includegraphics[width=\nwidth\linewidth]{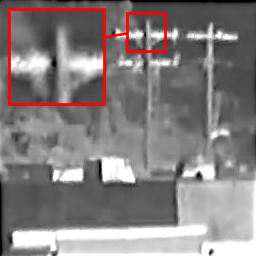} & \hspace{\nspace ex}\includegraphics[width=\nwidth\linewidth]{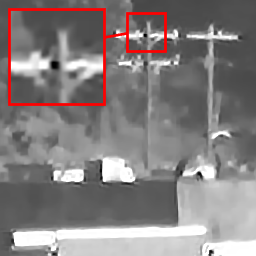} & \hspace{\nspace ex}\includegraphics[width=\nwidth\linewidth]{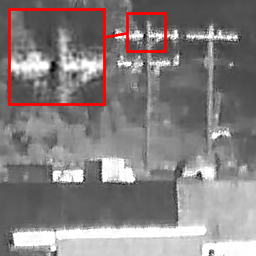} & \hspace{\nspace ex}\includegraphics[width=\nwidth\linewidth]{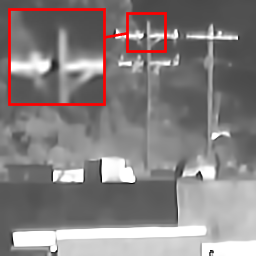} \\
	   (a) Observed & \hspace{-2ex} (b) Temp. Avg. & \hspace{-2ex} (c) SG\cite{Lou2013} & \hspace{-2ex} (d) NDL\cite{Milanfar2013} & \hspace{-2ex} (e) CLEAR\cite{Anantrasirichai2013} & \hspace{-2ex} (f) Proposed 
    \end{tabular}
    \vspace{-1ex}
    \caption{\textbf{Overall comparisons using real long-range static sequences}. The first sequence was obtained from Youtube \cite{youtube1}. The rest were captured by Panasonic Full HD Camcorder HC-V180K (aperture diameter 24mm and focal length 103mm), at a distance of 4km and temperature of 30$^\circ$C.}
    \label{fig:realexp}
\end{figure*}

\begin{figure*}[!]
    \centering
    \begin{tabular}{c c c c c}
      \includegraphics[width=0.18\linewidth]{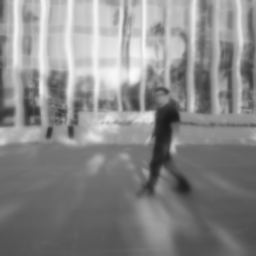} & \hspace{-2ex}\includegraphics[width=0.18\linewidth]{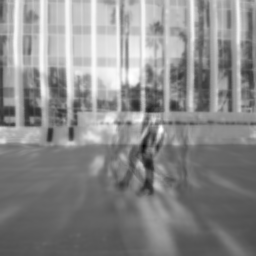} & \hspace{-2ex}\includegraphics[width=0.18\linewidth]{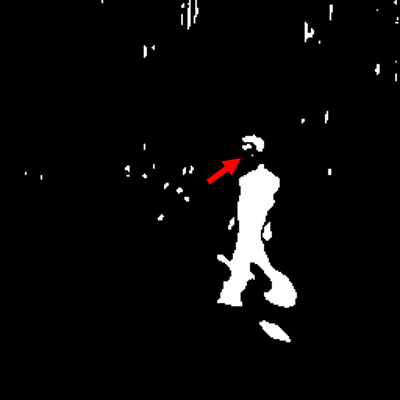} & \hspace{-2ex}\includegraphics[width=0.18\linewidth]{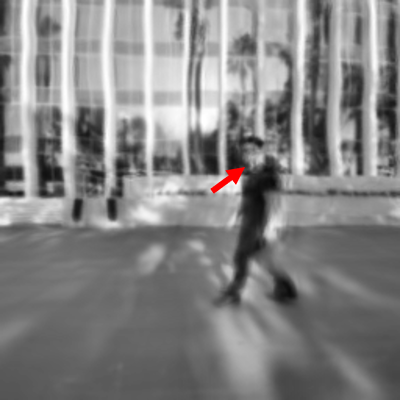} &
      \hspace{-2ex}\includegraphics[width=0.18\linewidth]{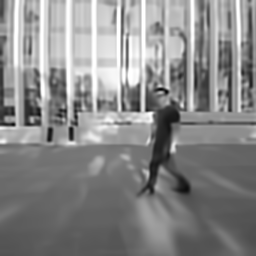}
	   \\
	   \includegraphics[width=0.18\linewidth]{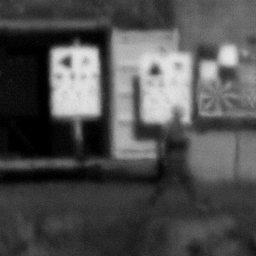} & \hspace{-2ex}\includegraphics[width=0.18\linewidth]{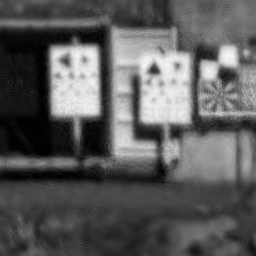} & \hspace{-2ex}\includegraphics[width=0.18\linewidth]{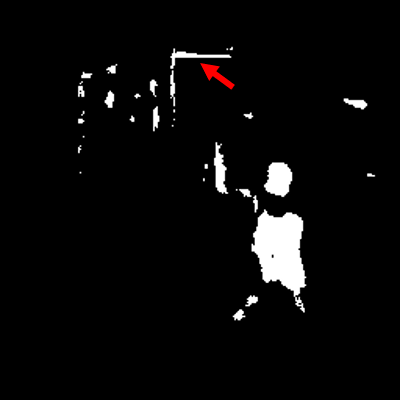} & \hspace{-2ex}\includegraphics[width=0.18\linewidth]{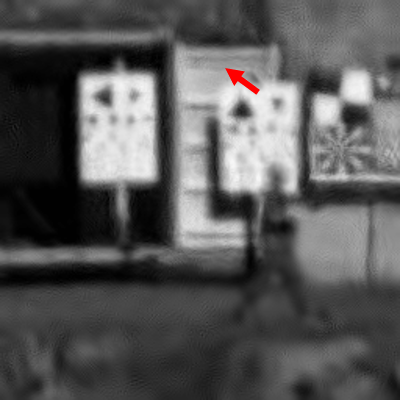} & \hspace{-2ex}\includegraphics[width=0.18\linewidth]{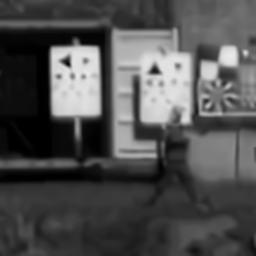} \\
	   (a) Input Frame & \hspace{-2ex} (b) CLEAR \cite{Anantrasirichai2013} & \hspace{-2ex} (c) Mask by \cite{Anan2018}& \hspace{-2ex} (d) Result by \cite{Anan2018} & \hspace{-2ex} (e) Proposed
    \end{tabular}
    \vspace{-1ex}
    \caption{\textbf{Overall comparison with synthetic and real dynamic sequences}. The first video is synthetically generated from clean video using the simulator presented in \cite{Our_paper}, whereas the last one is a real turbulence sequence from the NATO dataset. }
    \label{fig:movexp}
\end{figure*}

\begin{figure*}[!]
    \centering
    \begin{tabular}{c c c c c}
      \includegraphics[width=0.18\linewidth]{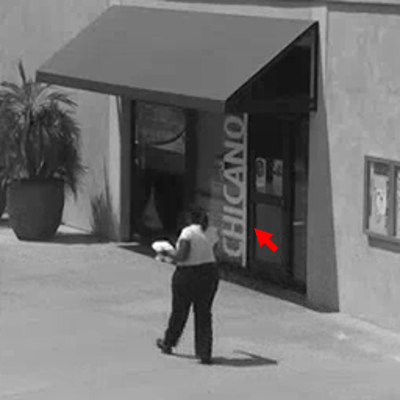} & \hspace{-2ex}\includegraphics[width=0.18\linewidth]{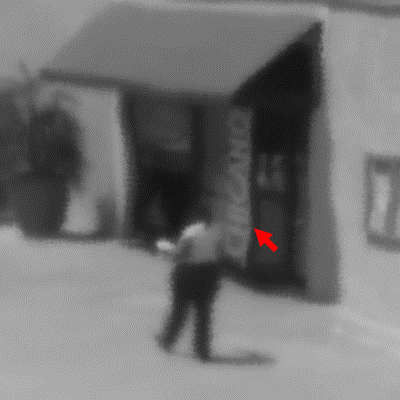} & 
      \hspace{-2ex}\includegraphics[width=0.18\linewidth]{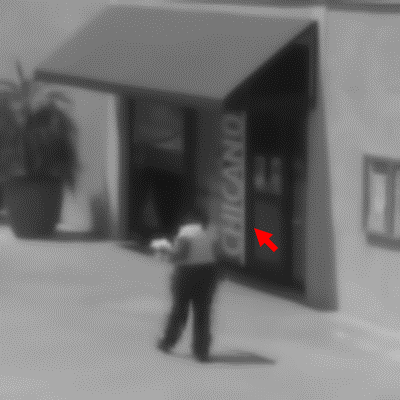} & 
      \hspace{-2ex}\includegraphics[width=0.18\linewidth]{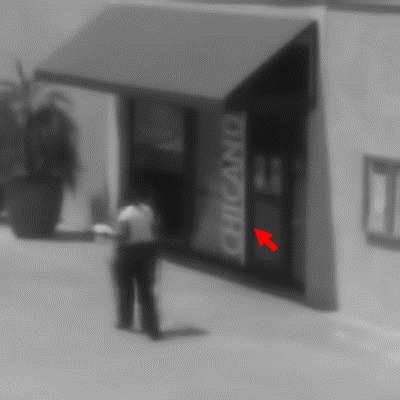} &
      \hspace{-2ex}\includegraphics[width=0.18\linewidth]{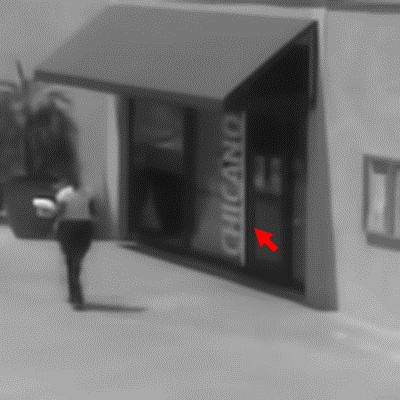}
	   \\
	   \includegraphics[width=0.18\linewidth]{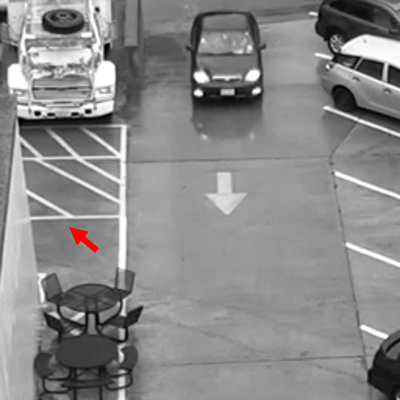} & \hspace{-2ex}\includegraphics[width=0.18\linewidth]{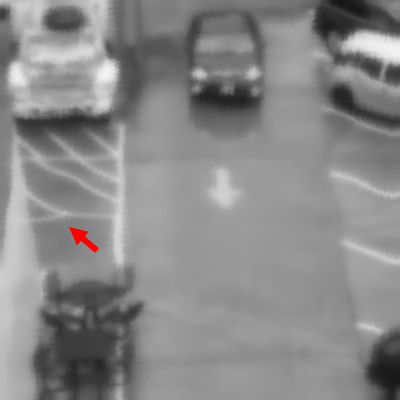} & 
      \hspace{-2ex}\includegraphics[width=0.18\linewidth]{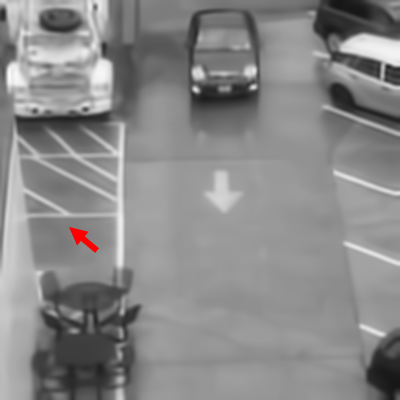} & 
      \hspace{-2ex}\includegraphics[width=0.18\linewidth]{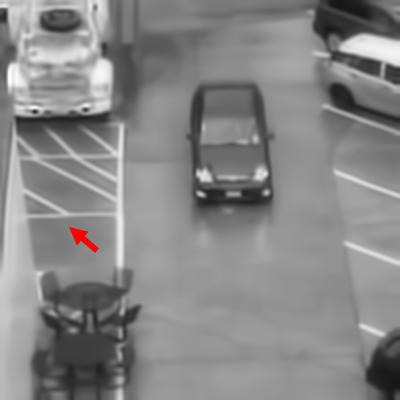} &
      \hspace{-2ex}\includegraphics[width=0.18\linewidth]{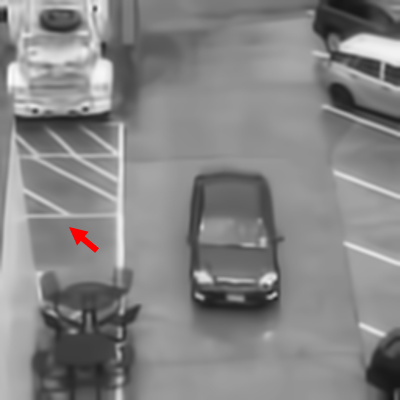}
	   \\
	   (a) Ground Truth & 
	   \hspace{-2ex} (b) Input frame \#10 & 
	   \hspace{-2ex} (c) Ours frame \#10  &
	   \hspace{-2ex} (d) Ours frame \#40 & 
	   \hspace{-2ex} (e) Ours frame \#70
    \end{tabular}
    \vspace{-1ex}
    \caption{\textbf{Additional results on synthetic dynamic sequences.} The video sequences are synthetically generated from clean videos from \cite{VIRAT} using simulator presented in \cite{Our_paper}. Note the consistent static background and well-maintained moving objects across frames. }
    \label{fig:AdditionalMov}
\end{figure*}

\textbf{Real sequences}. We compare the proposed method with existing methods on two sets of real turbulence sequences: Hot-air turbulence shown in \fref{fig:realexp1}, and a real long-range turbulence shown in \fref{fig:realexp}. We include both comparisons to demonstrate the universality of the proposed method. As we can see in both figures, the proposed method has more superior performance than the competing methods, both in terms of sharpness and visual consistency.

\textbf{Moving object sequences}. We finally compare methods for sequences containing moving objects. In particular, we compare our method with a very recent approach by Anantrasirichai et al. \cite{Anan2018}, which is a generalization of the wavelet fusion previously presented in \cite{Anantrasirichai2013} to moving sequences. The dataset we used consists of two synthetic video sequences and a real video sequence from the NATO RTO SET/RTG-40 dataset \cite{Leonard_Howe_Oxford}. 

The results shown in \fref{fig:movexp} suggest that static methods such as CLEAR \cite{Anantrasirichai2013} does not handle motion as it creates ghosting artifacts. If the object is moving quickly, e.g., the last row of \fref{fig:movexp}, the method will completely wash out the object. Segmentation-based method such as \cite{Anan2018} rely heavily on the quality of the masks, which are generally of sub-optimal quality. The proposed method demonstrates the most reliable reconstruction among the three, as is evident from the figure. Additional reconstruction results of the proposed method are shown in \fref{fig:AdditionalMov}.

\section{Conclusion}
We presented a unified turbulence mitigation algorithm that can restore both static and dynamic scenes. The method includes three new ideas: (i) A novel space-time non-local averaging method that can extract stable reference frames for static and dynamic scenes. (ii) A new lucky region fusion method that is robust against motion registration error. (iii) A new prior model for turbulent PSFs by utilizing the Zernike representation of the phase. Ablation studies of the method show that all components are essential for the overall performance. In particular, the reference extraction step has substantially better results than temporal averaging and robust principal component analysis, and the blind deconvolution outperforms existing deep learning based methods. We evaluated the algorithm using both synthetic and real turbulence image data. Compared to existing methods, our approach offers substantially better reconstruction quality on both static and dynamic scenes.

\section*{Acknowledgment}
The work is funded, in part, by the Air Force Research Lab and Leidos, and by the National
Science Foundation under grants CCF-1763896 and CCF-1718007. The authors would like to thank Michael Rucci, Barry Karch, Daniel LeMaster and Edward Hovenac of the Air Force Research Lab for many insightful discussions. The authors also thank Nantheera Anantrasirichai, a co-author of \cite{Anantrasirichai2013}, and Peyman Milanfar, a co-author of \cite{Milanfar2013}, for generously sharing their MATLAB code.

This work has been cleared for public release carrying the approval number 88ABW-2020-0292.

\ifCLASSOPTIONcaptionsoff
  \newpage
\fi

\bibliographystyle{IEEEtran}
\bibliography{egbibnew}

\end{document}